\documentclass[usenatbib]{mnras}

\usepackage{newtxtext,newtxmath}

\usepackage[T1]{fontenc}

\DeclareRobustCommand{\VAN}[3]{#2}
\let\VANthebibliography\thebibliography
\def\thebibliography{\DeclareRobustCommand{\VAN}[3]{##3}\VANthebibliography}

\usepackage{caption}
\usepackage{subcaption}
\usepackage{graphicx}	
\usepackage{amsmath}	

\usepackage{amssymb}	
\usepackage{xspace}
\usepackage{hyperref}

\newcommand{\Abacus}{\textsc{Abacus }\xspace}

\title[Accuracy of power spectrum]{Accuracy of power spectra in dissipationless cosmological simulations}

\author[S. Maleubre et al.]{
Sara Maleubre,$^{1}$\thanks{E-mail: sara.maleubre@lpnhe.in2p3.fr}
Daniel Eisenstein,$^{2}$
Lehman H. Garrison,$^{3}$
and Michael Joyce$^{1}$ \\
$^{1}$ Laboratoire de Physique Nucléaire et de Hautes Énergies, UPMC IN2P3 CNRS UMR 7585, \\ Sorbonne Université, 4, place Jussieu, 75252 Paris Cedex 05, France \\
$^{2}$ Harvard $\&$ Smithsonian Center for Astrophysics, 60 Garden St, Cambridge, MA 02138  \\
$^{3}$ Center for Computational Astrophysics, Flatiron Institute, 162 Fifth Ave., New York, NY 10010 \\
}

\date{Accepted XXX. Received YYY; in original form ZZZ}

\pubyear{2020}

\begin{document}
\label{firstpage}
\pagerange{\pageref{firstpage}--\pageref{lastpage}}
\maketitle

\begin{abstract}
We exploit a suite of large \emph{N}-body simulations (up to N=$4096^3$) performed with \Abacus, of scale-free models with a range of spectral indices $n$, to better understand and quantify convergence of the matter power spectrum. Using self-similarity to identify converged regions,
we show that the maximal wavenumber resolved at a given level of accuracy increases monotonically as a function of 
time. At the 1\% level it starts at early times from a fraction of $k_\Lambda$, the Nyquist wavenumber of the initial grid, and
reaches at most, if the force softening is sufficiently small, 
$\sim 2-3 k_\Lambda$ at the very latest times we evolve to. At 
the $5\%$ level, accuracy extends up to wavenumbers of order $5k_\Lambda$
at late times. Expressed as a suitable function of the scale-factor, accuracy shows a very simple $n$-dependence, allowing a extrapolation to place conservative bounds on the accuracy of \emph{N}-body simulations of non-scale free models like LCDM. We note that deviations due to discretization in the converged range are not well modelled by shot noise, and subtracting it in fact degrades accuracy.
Quantitatively our findings are broadly in line with the conservative assumptions about resolution adopted by recent studies using large cosmological simulations (e.g. Euclid Flagship) aiming to constrain the mildly non-linear regime.
On the other hand, we remark that conclusions about small scale clustering (e.g. concerning the validity of stable clustering) obtained using PS data at wavenumbers larger than a few $k_\Lambda$ may 
need revision in light of our convergence analysis.

 \end{abstract}




\section{Introduction}\label{sec:Intro}
The power spectrum (PS), together with its Fourier transform, the two-point correlation function (2PCF), are the most basic statistical tools employed to characterize clustering at large scales in cosmology. Building a precise theoretical framework for their calculation is crucial in order to fully exploit observational data coming from the next generation surveys, such as DESI \citep{DESI}, Vera C. Rubin Observatory LSST \citep{LSST} or Euclid \citep{Euclid}, that will open a new window in the era of ``precision cosmology''. In this context the nonlinear regime of gravitational evolution is of particular importance, as it will be a key to distinguishing among the plethora of exotic dark energy and modified gravity models \citep{MGnonlinear} as well as tightly constraining the LCDM scenario. Studies have estimated that to exploit fully the observed data, the matter PS in the range of scale $(0.1 \lesssim k/h \text{Mpc}^{-1} \lesssim 10)$ needs to be determined to   $1-2\%$ accuracy, depending on the specifications of the survey \citep{Huterer2005, Kitching2010, Hearin2011}. 
In the context of the preparation for the Euclid mission, a new study has been carried out with the goal of achieving a $1\%$ precision at non-linear scales \citep{EuclidPS2018}, which adds to the efforts of other large-volume simulations \citep{Ishiyama2020, Heitmann2020, Angulo2020}, the amendments to the widely used HaloFit Model in \citet{Takahashi2012}, or the most recent updated version of HMcode \citep{Mead2021}. 

In practice, calculation of predictions at these scales rely entirely on numerical simulations that use the $N$-body method. One important and unresolved issue in this context is the accuracy limitations on such simulations arising from the fact that they approximate the evolution of the dark matter phase space distribution using a finite particle sampling, as well as a regularisation at small scales of the gravitational force.  Despite the extensive use and spectacular development of $N$-body cosmological simulations over the last several decades, no clear consensus exists in the literature about how achieved accuracy, even for the PS, depends on the relevant parameters in an $N$-body simulation. Various studies have led to different conclusions (see e.g. \cite{splinter_1998, knebe_etal_2000, romeo_etal_2008, discreteness3_mjbm}), and in practice very different assumptions are made by different simulators about the range of resolved wavenumbers. A crucial difficulty is that, strictly, attaining the physical limit requires extrapolating the number of particles to be so large that there are many particles inside the gravitational softening length. Such a regime is unattainable in practice as the softening itself is small in simulations resolving small scales. Alternatively convergence may be probed also by comparing between two or more codes to assess the accuracy of their results (see e.g. \citet{schneider_et_al_2015, Garrison2019}), but this establishes only a \emph{relative convergence} that can give confidence in the accuracy of the clustering calculation but does not take into account the effects of discretization and dependence on the $N$-body parameters. 

In this article we study the small scale resolution limits on the PS arising from finite particle density and gravitational softening. To do so we use and extend a methodology detailed in \citet{Joyce2020} (hereafter P1) which uses the property of self-similarity of ``scale-free'' cosmological models, with an input power-law spectrum of fluctuations $P(k) \propto k^n$
and an Einstein de Sitter expansion law, to derive resolution limits for the 2PCF. In \citet{Garrison2021} 
(hereafter P2) this analysis has been extended to determine how resolution depends in detail on the gravitational softening, and in particular on whether such softening is held constant in comoving or proper coordinates.  In \citet{Leroy2020} the method has been shown to be a powerful tool also to infer the accuracy of measurements of the halo mass function and halo 2PCFs. In our analysis here we change focus to the PS, determining small scale resolution limits for it and examine how they compare with the limits for the 2PCF. While these previous articles based their analysis on a suite of simulations of a single scale-free model, with $n=-2$, we extend to simulations of a range of indices. This allows us to address more precisely the question of extrapolation to non scale-free cosmologies like the standard LCDM model.
Furthermore, we are able to exploit a new set of much larger scale-free simulations ($N=4096^3$ rather than $N=1024^3$) that were produced on the Summit supercomputer based on the methodology of the {\sc AbacusSummit} suite \citep{Maksimova+2021}. These allow us in particular to identify more precisely the finite box size effects that are an important limiting factor for analysis of scale-free models with redder spectra. Our analysis shows that we can indeed obtain precise information about convergence and spatial resolution cut-offs for the PS as a function of time, and that a level of precision in line with the requirements of forthcoming observational programs may be achieved. 

The article is structured as follows. In the next section we recall briefly what scale-free cosmologies are and explain how their self-similar evolution is able to provide a method for determining the precision with which statistical quantities are measured in $N$-body simulations. In Section 3 we describe our numerical simulations and then detail our calculation of  the PS. In Section 4 we present and analyse our results, detailing the criteria used to identify converged values of the PS and temporal regions within which a given accuracy is attained, and using them to infer limits on resolution at small scales in the different models. Section 5 describes how the results can be extrapolated to infer conservative resolution limits in non scale-free (LCDM type) cosmologies. In the final section we summarise our results and discuss them in relation to both the related works P1 and P2, and more broadly to the literature on numerical study of the dark matter PS in cosmology.

\section{Scale-Free cosmologies and Self-similarity}\label{sec:SF_SS}
One of the limitations of using $N$-body simulations for the study of cosmological systems comes from its inherent discreteness, as the continuous phase space density describing dark matter is sampled using a finite number $N$ of objects. Such simulations thus introduce a set of \emph{unphysical} parameters which necessarily limit the range of time and length scales which they resolve. These parameters can be divided into the \emph{numerical parameters} controlling the approximations made in the integration of the $N$-body dynamics and forces, and the 
parameters introduced in passing from the physical model to the $N$-body system:
the mean interparticle spacing $\Lambda$, the force softening scale $\epsilon$ and the size of the periodic box $L$, and a starting redshift $z_i$ that can be parametrized by the value of $\sigma_i(\Lambda,zi)$, the square root of the variance of normalized linear mass fluctuations in a top-hat sphere of radius $\Lambda$. We will refer to the latter as the \emph{discretization parameters}. 

Following this distinction, we can divide the consideration of limitations on $N$-body simulations into two main parts: the issue of convergence of the numerical solution of a specific configuration (fixed $\Lambda$, $\epsilon$, L and $\sigma_i$), and the convergence towards the continuum cosmological model (extrapolated by taking the appropriate limit of the \emph{discretization parameters}). The former can be treated by studying stability under variations of \emph{numerical parameters} such as time-stepping or force accuracy. We will discuss this point only briefly here as it has already been treated extensively in P1 and P2. The latter is more challenging and, as we have discussed in the introduction, lacks a consensus in the literature. We will be focusing primarily on it here, considering convergence to the physical limit of the matter PS.

The value of scale-free models in the context of physical resolution of $N$-body simulations relies on the self-similarity of their evolution: \emph{temporal evolution of clustering is equivalent to a well defined rescaling of the spatial coordinates}. This is the case because such models are characterised by just one length scale and one time scale: their initial (linear) PS of fluctuations is a simple power law $P(k)\propto k^{n}$, and an Einstein-de-Sitter expansion law $a\propto t^{2/3}$ (where $a$ is the scale factor). The single length scale
can thus be defined as the non-linearity scale $R_{NL}$ given by
\begin{equation}
    \sigma^2_{lin}(R_{NL},a)=1 \xrightarrow{\text{linear theory}} R_{NL}\propto a^{\frac{2}{3+n}}
    \label{eq:RNL}
\end{equation}
where $\sigma^2_{lin}$ as the variance of normalized linear mass fluctuations in a sphere, while the time scale 
is fixed by the normalization of the Hubble law (i.e. by the mean mass density combined with Newton's constant $G$). For the case of statistics such as the PS which are a function of wavenumber $k$ and time,
it follows simply by dimensional analysis that a suitable dimensionless definition of the 
statistic $f$  can be written as a function of  $k\,R_{NL}(a_0)$
where $a_0$ is some reference scale-factor: 
\begin{equation}
    f(k,a)=f(k\,R_{NL}(a_0),a/a_0)\,.
\end{equation}
As the reference scale-factor is itself arbitrary (because of the EdS expansion law) we can take $a_0=a$, and obtain 
\begin{equation}
f(k,a)=f_0(k\,R_{NL}(a))
\end{equation}
where $f_0$ is independent of time. For the case of the PS we use in our analysis the canonical 
definition of the dimensionless PS  
\begin{equation}\label{eq:dimensionlessPS-def}
    \Delta^{2}(k,a)=\frac{k^3 P(k,a)}{2\pi^2}\,, 
\end{equation}
and thus self-similar behaviour corresponds to
\begin{equation}\label{eq:dimensionlessPS-scaling}
    \Delta^{2}(k,a)=\Delta_0^2(k\,R_{NL}(a))
\end{equation}
In $N$-body simulations of scale-free cosmologies, any deviations from this self-similar evolution can only be due to  
\emph{unphysical scales}. Conversely results can represent the PS in the desired  physical limit --- which must be independent of these parameters --- only to the extent that the rescaled dimensionless PS statistic becomes independent of time. As in P1 and P2 we caveat that self-similarity does not in itself \emph{prove} definitively 
that a measured PS represents the physical limit, as it is not impossible that an unphysical parameter may 
itself have a self-similar scaling. In particular, as highlighted in P1 and P2, we need to be careful 
to ensure that there are not hidden errors due to time-stepping, whose errors seem approximately self-similar even when the time step size is not an equispacing of $\log a$.

\section{Numerical simulations}\label{sec:NumSim}
\subsection{Abacus code and simulation parameters}

We have performed simulations using the \Abacus $N$-body code \citep{Garrison+2021}.  \Abacus is designed for high-accuracy, high-performance cosmological $N$-body simulations, exploiting a high-order multipole method for the far-field force evaluation and GPU-accelerated pairwise evaluation for the near-field.  The larger, $N=4096^3$ simulations in this work were run as part of the \textsc{AbacusSummit} project \citep{Maksimova+2021} using the Summit supercomputer of the Oak Ridge Leadership Computing Facility.

We report results here based on the simulations listed in \autoref{tab:Simulations}. We have simulated three different exponents for the PS ($n=-1.5,-2.0,-2.25$) in order to probe the range of exponents relevant to structure  formation in a LCDM cosmology. Ideally we would extend this range further towards $n=-3$, but as we will discuss below, $n=-2.25$ represents in practice the accessible limit below which 
even our largest simulations would be swamped by finite box size effects. Indeed our simulations include for each $n$ a pair of simulations 
with  $N=1024^3$ and $N=4096^3$ for which parameters are otherwise identical.

The remaining crucial parameter for our simulations is the softening length. Again the simulations in \autoref{tab:Simulations} have been chosen to include for each $n$ a least one pair of simulations (with $N=1024^3$) which differ only in this parameter (and for $n=-2.0$ a range of different softenings). In \Abacus the gravitational force is softened (see P2 for the explicit functional form and a more detailed discussion) from its Newtonian form using a compact spline  which reverts to the exact Newtonian force at $2.16 \epsilon$, where $\epsilon$ is defined for convenience as an ``effective Plummer smoothing'' i.e. the softening of a Plummer model (with two body force $F(r) \propto (r^2+\epsilon^2)^{-3/2}$) with the same minimal pairwise dynamical time. The values of $\epsilon$ in our simulations are given in \autoref{tab:Simulations} as a ratio to the mean inter-particle spacing $\Lambda$. For the values given without an asterisk this value is fixed throughout the simulation i.e. the smoothing length is fixed in comoving units. For the values with an asterisk the value given is that at the time of our first output, at the scale-factor $a=a_0$ defined below. For $a<a_0$ the smoothing is again fixed in comoving coordinate, while for  $a>a_0$ it is kept fixed in physical coordinates i.e. $\epsilon/\Lambda \propto 1/a$. Our choices for the simulations with $n\neq -2$ have been guided by the detailed study in P2 of the effect of softening on resolution, based on the study of the 2PCF in a very large suite of $n=-2$ simulations (including the subset of four $1024^3$ simulations of $n=-2.0$ we report here). 
We will see below that the qualitative and quantitative conclusions of P2 found using the 2PCF extend also to the PS. 

 \begin{table*}
  \begin{tabular}{lccccc}
    \hline
    Name & n & $N$ & $\epsilon/\Lambda$  & $S_{f}$  & $\log_{2}(a_f/a_0)$ \\
    \hline
    \hline
    N4096\_n1.5\_eps0.3phy & -1.5 & $4096^3$ & $0.3^{*}$ & 29 & 3.625 \\
    N1024\_n1.5\_eps0.3phy & -1.5 & $1024^3$ & $0.3^{*}$ & 29 & 3.625\\
    N1024\_n1.5\_eps0.03com & -1.5 & $1024^3$ & 1/30  & 29 & 3.625 \\
    \hline
    N4096\_n2.0\_eps0.3phy & -2.0 & $4096^3$ & $0.3^{*}$ & 35 & 2.917 \\
    N1024\_n2.0\_eps0.3phy & -2.0 & $1024^3$ & $0.3^{*}$ & 37 & 3.083 \\
    N1024\_n2.0\_eps0.03com & -2.0 & $1024^3$ & 1/30 & 37 & 3.083\\
    N1024\_n2.0\_eps0.02com & -2.0 & $1024^3$ & 1/60 & 37 & 3.083\\
    N1024\_n2.0\_eps0.07com & -2.0 & $1024^3$ & 1/15 & 37 & 3.083 \\
    \hline
    N4096\_n2.25\_eps0.3phy & -2.25& $4096^3$ & $0.3^{*}$ & 35 & 2.1875\\
    N1024\_n2.25\_eps0.3phy & -2.25& $1024^3$ & $0.3^{*}$ & 37 & 2.3125\\
    N1024\_n2.25\_eps0.03com & -2.25& $1024^3$ & 1/30 & 37 & 2.3125\\
    \hline
  \end{tabular}
  \caption{Summary of the $N$-body simulations used for the analysis in this paper. The first column shows the name of the simulation, $n$ is the spectral index of the initial PS, and $N$ the number of particles. The fourth column gives the ratio of the effective Plummer force smoothing length $\epsilon$ to  mean inter-particle separation (equal to the initial grid spacing $\Lambda$), for $a<a_0$ (as defined by \autoref{eq:def_a0}, the time of our first output). For the cases without an asterisk this is its value at all times (i.e. the smoothing is fixed in comoving coordinates) while for the cases with an asterisk the smoothing for $a>a_0$ is fixed in proper coordinates. The last two columns give, respectively, the value of the time parameter $S$ at the last snapshot and the final scale factor relative to that at first output. Note that given that the first output is at $S=0$, the number of outputs for each simulation is $S_f+1$. 
  }
  \label{tab:Simulations}
 \end{table*}

Initial conditions have been set up using the standard method based on the Zeldovich Approximation (ZA) but modified as described in detail in \cite{garrison_et_al_2016}, to include both next order 2LPT corrections to the ZA as well as corrections for deviations from the ZA due to discretization described analytically by the 
"particle linear theory" (PLT) derived in \citet{Joyce2005, marcos_et_al_PLT_2006}.  The initial conditions are then characterised by two parameters: $\sigma_i$, the value of the top-hat variance at the grid spacing in the initial configuration, and $a_{PLT}$, the target scale factor at which the PLT evolved modes coincide exactly with the ZA evolved modes prior to correction. For all our simulations here we have $\sigma_i=0.03$, and $a_{PLT}=a_0$ where $a_0$ corresponds to the time defined by 
\begin{equation}
    \label{eq:def_a0}
    \sigma_{lin}(\Lambda,a_0)=0.56
\end{equation}
i.e. to the epoch at which fluctuations of peak-height $\nu\approx3$ are expected to virialize in the spherical collapse model ($\sigma_{lin}\approx\delta_c/\nu$, with $\delta_c=1.68$), and the first non-linear structures appear in the simulations. 

The accuracy of the numerical integration of the initialized $N$-body system in \Abacus is determined in practice essentially by a single parameter, the time-step parameter $\eta$. For all simulations here we have used $\eta=0.15$, on the basis of the extensive tests of the convergence of the 2PCF reported in P1 and P2. We do not explore here further the sensitivity of results to the choices of $\sigma_i$ and $a_{PLT}$, based also on tests which have been performed in the context of the studies of P1 and P2. In practice our $\sigma_i$ is sufficiently small so that reducing it further leads to no significant change to the initial conditions, because of the application of the PLT corrections. The effects of the choice of the value of $a_{PLT}$ is a subject we will explore in future work. 

As in P1 and P2, we have saved the outputs (full particle configurations) of our simulations starting from $a=a_0$ and then at subsequent times separated by intervals in which the characteristic non-linear mass $M_{NL}$ grows by a factor of $\sqrt{2}$. Given that $M_{NL} \propto R_{NL}^3$, the scaling in \autoref{eq:RNL} implies that the outputs correspond to scale-factors $a_i$ where 
\begin{equation}
    \log_2\left(\frac{M_{NL}(a_{i+1})}{M_{NL}(a_{i})}\right)=\frac{1}{2}=\frac{6}{3+n}\log_{2}\left(\frac{a_{i+1}}{a_i}\right)\,.
\end{equation}
It is convenient then to define the time variable $S$ by 
\begin{equation}\label{eq:Sdef}
    S=\frac{12}{3+n}\log_2\left(\frac{a_s}{a_0}\right)
\end{equation}
with the outputs corresponding to $S=0,1,2 \dots$.
The final time $S_f$ up to which we have integrated is dictated by two competing considerations. On the one there is a limitation on numerical cost arising from the use of a global time-step in \Abacus which 
means that it can no longer integrate efficiently when the central densities of halos become too large. 
On the other hand finite box size effects become dominant at sufficiently long times. For $n=-1.5$, 
it is the former limitation which dictates the 
stopping time, while for $n=-2.25$ it is the latter.
This will be illustrated explicitly in our analysis below. 


\subsection{Power spectrum calculation}\label{sec:PScalculation}

In this section we describe the method we have used to measure the matter PS of our particle data sets.  

We recall first that the PS $P^d (k)$ of a distribution of discrete equal mass points, in a periodic cube of side $L$, may be calculated, for wave-vectors $\mathbf{k}=(2\pi/L) \mathbf{n}$ 
where $\mathbf{n}$ is any non-zero triplet of integers, as 
\begin{equation}
    P^d (k)\equiv \frac{|\delta^d(\mathbf{k})|^2}{L^3}
\end{equation}
where 
\begin{equation}
\label{eq:delta-def}
    \delta^d(\mathbf{k})=\frac{L^3}{N}\sum_j  e^{i\mathbf{r}_j\cdot\mathbf{k}}
\end{equation}
being the sum over all $N$ particles in the volume
(with $\delta^d(\mathbf{k})$ the coefficients of the Fourier sum for the periodic system).

Because of the excessive computational cost of performing directly the summation in \autoref{eq:delta-def}, we use instead, as is commonplace, an approximation which replaces the Fourier sum by an FFT on a grid of $N_g$ points covering the same volume. We thus assign a fraction of each particle's mass to each cell to obtain the number density in a cell $klm$ given, in units of the cell volume, as 
\begin{equation}
    n_{klm}=\sum_{i=1}^{N}W(\mathbf{r}_i)
\end{equation}
where $W(\mathbf{r}_i)$ is the mass assignment function, with $\mathbf{r}_i$ the relative position of the particle in the grid and the centre of the cell. We have used a so-called Triangular Shape Cloud (TSC) mass assignment, of which the functional form is given in Appendix~\ref{appendix}. 
The relation between the PS $P^f$ estimated on the FFT grid and the exact result is then \citep[see][]{Book, Jing}
\begin{equation}
\label{eq:aliasing}
    P^f(\mathbf{k})=  \sum_{\mathbf{n}}|W(\mathbf{k}+2k_{g}\mathbf{n})|^2P^d(\mathbf{k}+2k_{g}\mathbf{n}) 
\end{equation}
where $W(\mathbf{k})$ is the FT of the window function $W(\mathbf{r})$, $k_{g}=\pi N_g^{1/3}/L$ is the Nyquist frequency of the FFT grid, and the summation is over all triplets of integers $\mathbf{n}$ (including $\mathbf{n}=\mathbf{0}$).

\autoref{eq:aliasing} tells us how the use of the FFT grid limits the accuracy of our measurement of the PS,  due to ``aliasing" which becomes important for wavenumbers of modulus $k$ approaching $k_g$. As detailed further in Appendix \ref{appendix}, we apply a commonly used correction which reduces these effects  dividing $P^{f}$ by 
 $\sum_n W^2(\mathbf{k}+2k_g\mathbf{n})$ and cutting the results at $k_g/2$, where the correction is no longer valid (see \citet{Angulo2007,Sato2011,Takahashi2012}). 
 
 We need for our study to obtain highly accurate PS measurements at scales well below the Nyquist frequency of the \emph{initial particle grid}, 
 $k_\Lambda= \pi N^{1/3}/L $. Our analysis establishes a posteriori what the maximal required wavenumber 
 is, and what level of accuracy we need. In 
 practice the calculation times for the PS of
 the full particle loads of even our smaller
 $1024^3$ simulations becomes excessive for 
 $N_g \sim 3000^3$, which does not give sufficient resolution 
 for our requirements. To overcome this 
 limitation we employ the widely used technique of ``folding" (see e.g. \citet{Jenkins1998}): the PS is calculated (using the FFT method) on a new particle configuration obtained by superimposing sub-cubes of the initial configuration i.e. the particle positions are modified by taking  $\mathbf{x}\rightarrow\mathbf{x}\%\left(L/2^m\right)$, where the operation $a\%b$ stands for the reminder of the division of \emph{a} by \emph{b} and $m$ is a positive integer. It can be shown easily that the PS of the two configurations are identical for the modes which are common to them. If one uses the same FFT grid, for  
 each folding we can access accurately a maximal wavenumber which is multiplied by two, at the cost of losing half the wave-vectors (as they are integer multiples of $2^{m+1}\pi/L$).   
 
 For the results reported here we have used a grid with $N_g=1024^3$. For the $1024^3$ simulations we have calculated the PS using the full particle loads 
 for foldings $m={0,1,2,3,4,6}$; for 
 the $4096^3$ simulations we have used the available random down-sampling to $10\%$ of the full particle loads,
 and have included an additional folding with $m=8$ (in order to reach the same maximal wavenumber in units of $k_\Lambda$).

 The next step is to use our measured $P^f$ to estimate the PS of the matter field corresponding to an ensemble average over realizations of the theoretical model. Given that the latter is a function only of $k$, we use, as is standard, an estimator for the average in bins defined
 by spherical shells, taking
\begin{equation}
\label{eq:PS-estimator}
    P(k; \Delta k)=\left(\frac{1}{N_k}\sum_{\mathbf{k} \in[k, k+\Delta k]} P^f(\mathbf{k})\right)-\frac{L^3}{N} \left(\frac{1}{f}-1\right)
\end{equation}
where the sum in the first term is over the $N_k$ modes with modulus in the range $[k, k+\Delta k]$, and the second term is a correction for the shot noise added to the PS of the particle distributions
when a (random) down-sampling to a fraction $f$ of the total particle number is applied. We have calculated using bins with constant logarithmic spacing, taking twelve bins per octave, i.e.  $1+(\Delta k/k)=2^{1/12}$. This choice of binning 
is convenient because it ensures that the bins for
different snapshots still match when rescaled by $R_{NL}$ (which changes between snapshots by multiples of $2^{1/6}$). In other words our binning is designed so that it doesn't break self-similarity. In the results presented below, in order to reduce statistical noise sufficiently, we have rebinned by grouping four such bins, corresponding to $\Delta k/k \approx 0.26$. 

The shot noise subtraction in \autoref{eq:PS-estimator} accounts for (random) down-sampling, and vanishes when $f=1$ (as in all of our $1024^3$ simulations). We note that this 
formula does not include subtraction of the shot noise term characteristic of the PS (of any stochastic point process) at asymptotically large $k$ (here $P(k\rightarrow \infty)=L^3/N=\Lambda^3$). Some authors advocate making such a subtraction to correct for discreteness effects, while others argue against doing so (see e.g. \citet{heitmann_et_al_part1}). 
Our method here is precisely designed to detect how $N$-body discretisation modifies the continuum PS, and so we do not wish to make any a priori assumption about how to model such effects. It is nevertheless an interesting question to assess specifically whether shot noise subtraction can
be effectively used to correct for discreteness effects. We return to a discussion of this issue therefore after our main analysis, in \autoref{sec:PoisonNoise}.
As detailed there, it turns out that shot noise is not a good model in the range of wave-number where we can establish accurate convergence to 
the physical limit. 

Finally we combine, for each snapshot, the measurements with different foldings to obtain a single estimate for each bin. Since the ratio of a given wavenumber $k$  to the Nyquist wavenumber of the grid scales as $1/2^m$, comparison of the different measurements of
$P(k; \Delta k)$ in a given simulation allows us to assess the magnitude of the (systematic) inaccuracies arising from the FFT grid. Doing so we have concluded that, for $k< k_g/2$, the accuracy of our PS measurement is always better than $0.2\%$. 
As this level of accuracy is sufficient for our purposes here, we construct a single measurement of $P(k; \Delta k)$ for each snapshot taking for each bin the measurement from the available folding with the lowest value of $m$.  

For economy of notation we will omit everywhere the bin width and label bins by a $k$ corresponding to their \emph{geometric centre}. Likewise we calculate $\Delta^2(k)$ using \autoref{eq:dimensionlessPS-def} 
at this same value for $k$.

\section{Results}\label{sec:Results}
\subsection{Self-similarity of the dimensionless power spectrum}

\begin{figure*}
  \begin{subfigure}[b]{\textwidth}
    \centering
    \includegraphics[width=1.\linewidth]{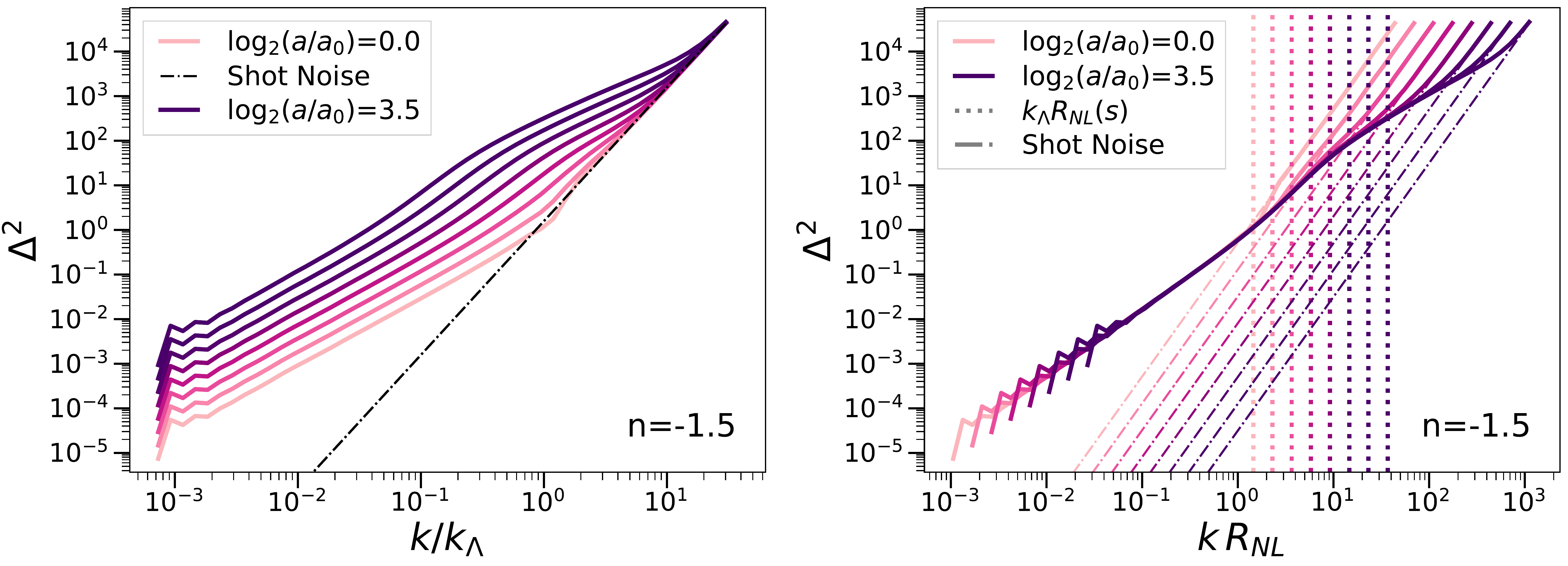}
  \end{subfigure}
  \begin{subfigure}[b]{\textwidth}
    \centering
    \includegraphics[width=1.\linewidth]{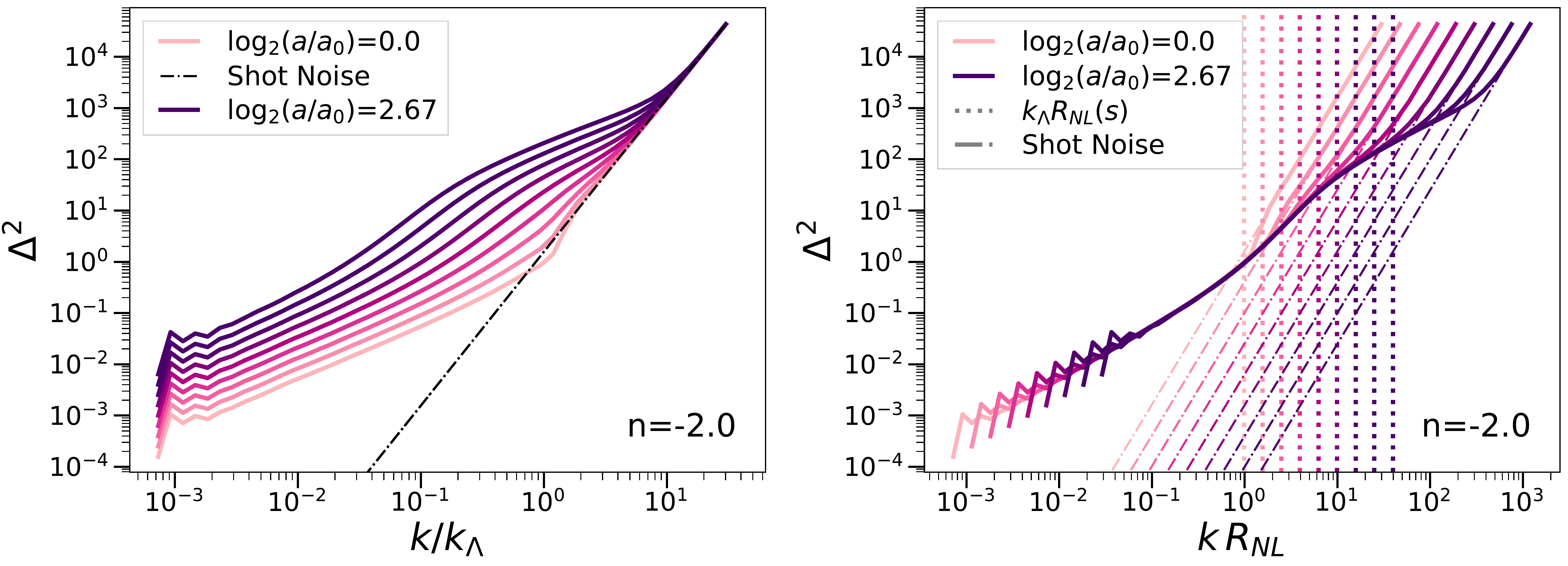}
  \end{subfigure}
  \begin{subfigure}[b]{\textwidth}
    \centering
    \includegraphics[width=1.\linewidth]{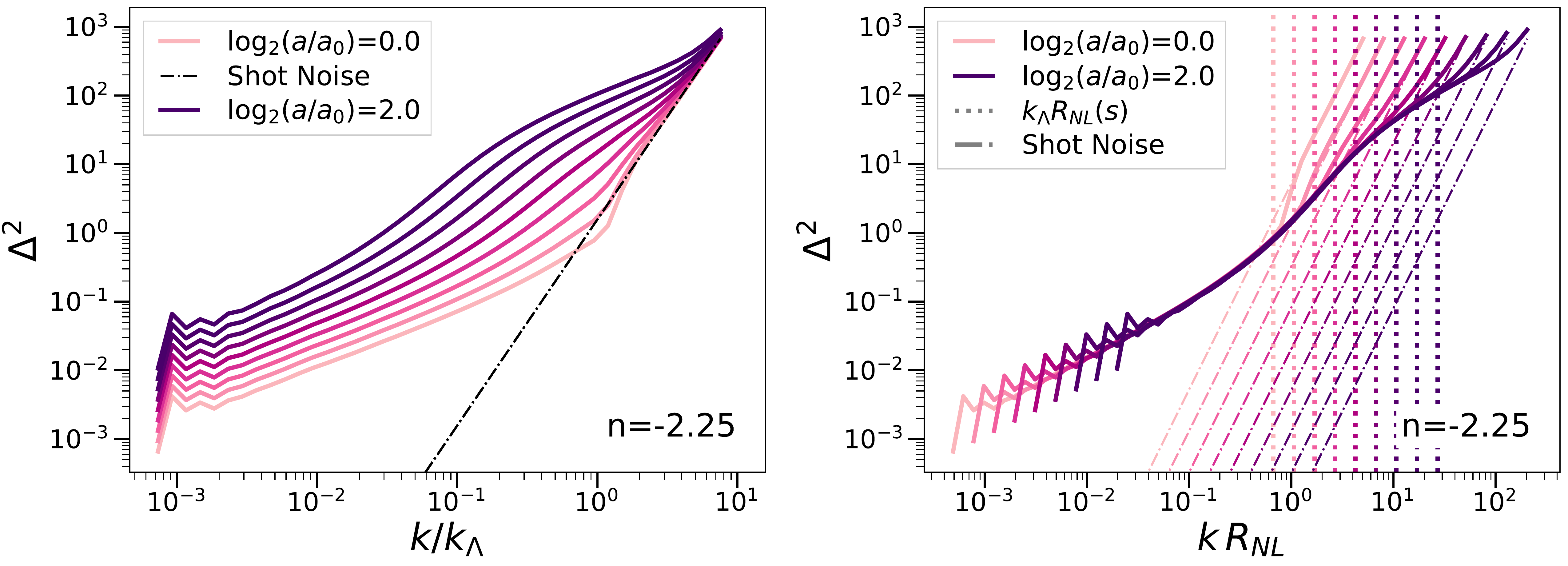}
  \end{subfigure}
  \caption{Dimensionless PS $\Delta^{2}$ as a function of wavenumber (leftmost panels) and of rescaled wavenumber (second column) of our largest simulations ($N=4096^3$), for which self-similar evolution corresponds to a $\Delta^{2}$ which is invariant in time. The times shown correspond to every fourth snapshot $S=0,3,7\cdots $ (where $S$ is as defined in \autoref{eq:Sdef}) over the total time-span of the simulations. The vertical dotted lines in the right panels correspond to the rescaled Nyquist wavenumbers for each displayed epoch. The shot noise limit, $\Delta^2_{\text{shot}}=(\pi/2)(k/k_{\Lambda})^3$, is represented with dash-dotted lines. The right panels show the appropriately shifted line for each displayed epoch.}
  \label{fig:PS_selfsim}
\end{figure*}

As we have discussed, in scale-free cosmologies independence of the discretization parameters 
in an $N$-body simulation should correspond to a self-similar evolution for the PS, which is 
conveniently characterised in terms of the dimensionless $\Delta^2$ 
as given in \autoref{eq:dimensionlessPS-scaling}. \autoref{fig:PS_selfsim} shows the evolution of $\Delta^2$ as a function of time (parametrised by the variable $\log_2(a/a_0)$) for each of the three different spectral indices we have simulated, in all cases for the largest simulation (with $N=4096^3$) in \autoref{tab:Simulations}.
For each index there are two panels: the left panel gives $\Delta^2(k)$ as a function of $k/k_\Lambda$ 
(where $k_\Lambda$ is the Nyquist wavenumber of the initial particle grid), while the right panels give the same rescaled quantity as a function of the variable $k R_{NL}$. Thus self-similar evolution corresponds to the superposition in this latter plot of the data at different times. In the right panel vertical dotted lines are plotted of the rescaled $k_\Lambda R_{NL}$ for each epoch, while dotted-point lines in both panels represent the shot-noise $L^3/N$.

These plots illustrate qualitatively the crucial features we will analyse more quantitatively below.
For all spectral indices, we can see the same basic trends: self-similarity --- identified as proximity to the common locus traced by the different curves --- propagates monotonically to larger $kR_{NL}$ (and thus larger $\Delta^2$) as a function of time. At early times the self-similar range is restricted to linear scales ($\Delta^2 \ll 1$), but it develops progressively in the non-linear regime ($\Delta^2 \gg 1$). At each time we can identify an apparent upper cut-off (in k-space) to self-similarity where it branches off from the common locus. As a function of time this cut-off starts very close to $k_\Lambda$, this corresponds to the imprint of the initial lattice, as in the initial conditions the input PS is represented only up to this wavenumber. As non-linear structure 
develops at smaller scales, this cut-off appears to progressively overtake the wavenumber $k_\Lambda$ and at the final time appears to be significantly larger. Regarding the shot noise, we can see two main behaviours: at early times evolution is still self-similar for values of the PS larger than a rescaled $L^3/N$ term, while for later times self-similarity is broken at a smaller k and it's only at large k when $\Delta^2$ is completely dominated by it. 

Differences between the evolution for the three cases, which we will see in greater detail below, are also evident on visual inspection. Firstly the range of scale, and the maximal value of $\Delta^2$, for which there is self-similarity decreases markedly from $n=-1.5$ to $n=-2.25$. This is just a reflection of the smaller range of scale-factor which is accessible in simulations of a fixed size. Furthermore we can detect that the quality of the superposition is significantly poorer for $n=-2.25$. This is the result of the very much more significant finite size effects present for this very red index: in fact, as we will see below, the quality of the self-similarity is so much less good in this case that we can exploit these simulations to obtain information about the propagation of resolution at the $1\%$ level only in a very limited range.


\subsection{Convergence to self-similarity}

\begin{figure*}
  \begin{subfigure}[b]{0.49\textwidth}
    \centering
    \includegraphics[width=\textwidth]{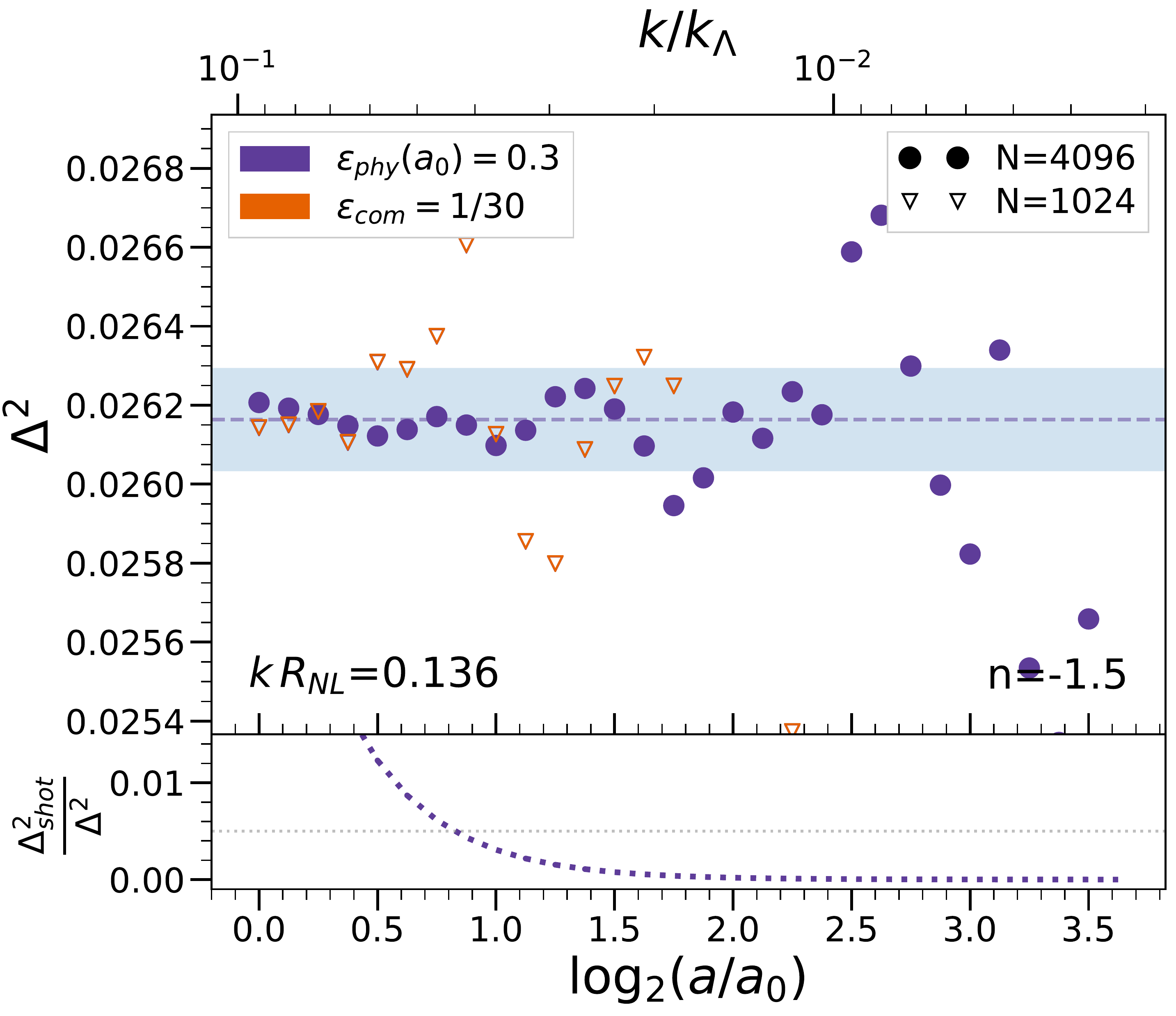}
  \end{subfigure}
  \begin{subfigure}[b]{0.49\textwidth}
    \centering
    \includegraphics[width=\textwidth]{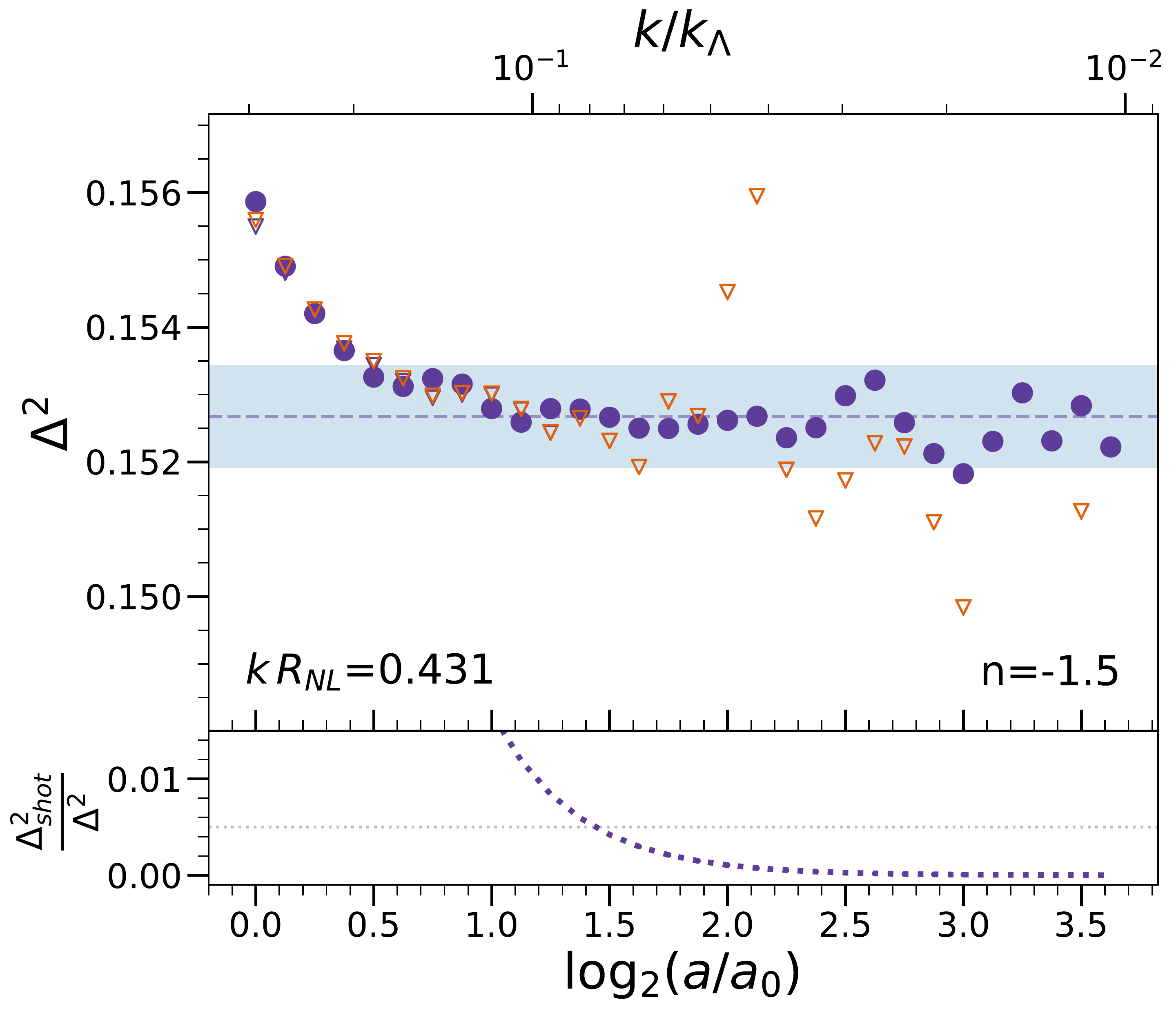}
  \end{subfigure}    
  \begin{subfigure}[b]{0.49\textwidth}
    \centering
    \includegraphics[width=\textwidth]{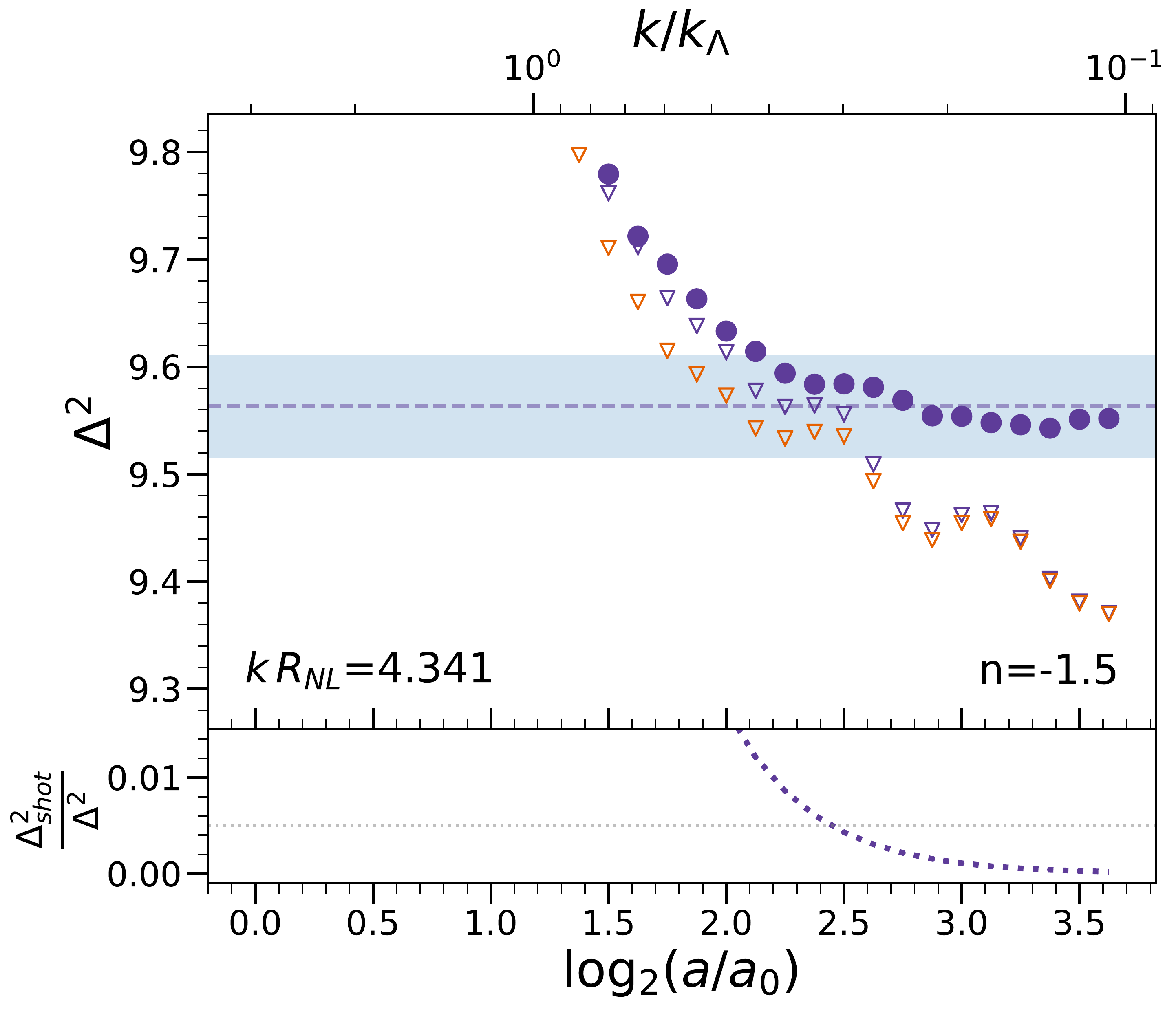}
  \end{subfigure}
  \begin{subfigure}[b]{0.49\textwidth}
    \centering
    \includegraphics[width=\textwidth]{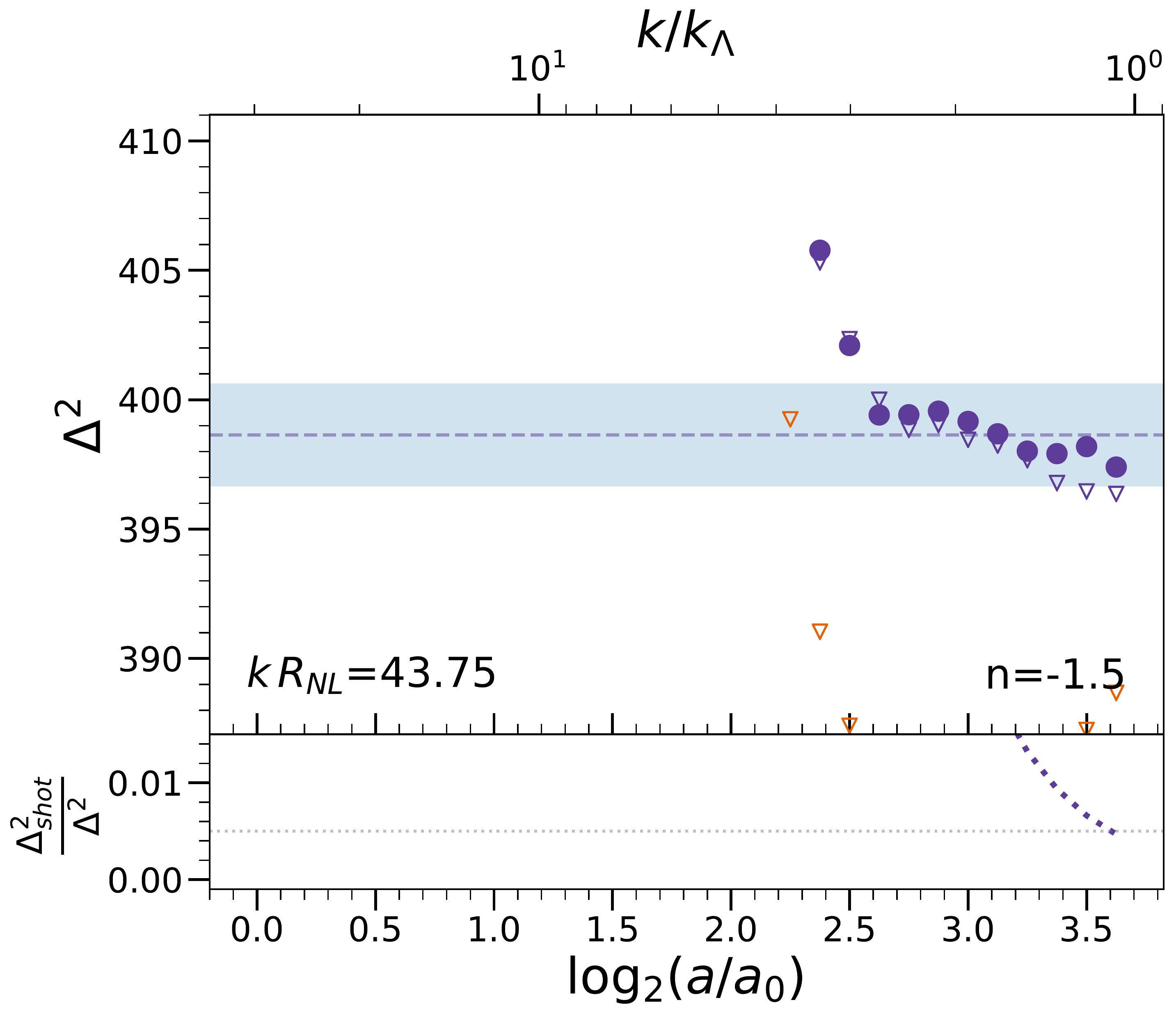}
  \end{subfigure}
  \caption{Evolution of the dimensionless PS as a function of logarithmic scale factor $\log_2(a/a_0)$ (lower $x$-axis) and as a function of $k/k_\Lambda$ (upper $x$-axis) for a set of given rescaled bins labelled by their bin value $kR_{NL}$, 
  in each of the indicated $n=-1.5$ simulations (cf. \autoref{tab:Simulations}). Circles correspond to the largest ($N=4096^3$) simulations and triangles to the $N=1024^3$ simulations. The simulations using physical softening are plotted in purple and those with comoving softening in orange. The horizontal dashed line marks our estimated converged value ($\Delta^2_{\text{conv}}$) in each bin, determined in the largest simulation as described in the text. The blue shaded region indicates that within $\pm 0.5\%$ of this value. These plots show in particular that the convergence of the PS depends on box-size for larger scales (i.e. small \emph{k}) 
  and on the force smoothing for smaller scales. The lower panel of each plot shows, for the larger simulation, the fraction of $\Delta^2$ represented by a shot-noise term 
  ($\Delta^2_{\text{shot}}=(\pi/2)(k/k_{\Lambda})^3$),
  and the dotted horizontal line marks $0.5\%$. We note that the observed deviations from the estimated converged value are not well approximated by such a term.}
  \label{fig:n15_convplots}
\end{figure*}

\begin{figure*}
  \begin{subfigure}[b]{0.49\textwidth}
    \centering
    \includegraphics[width=\textwidth]{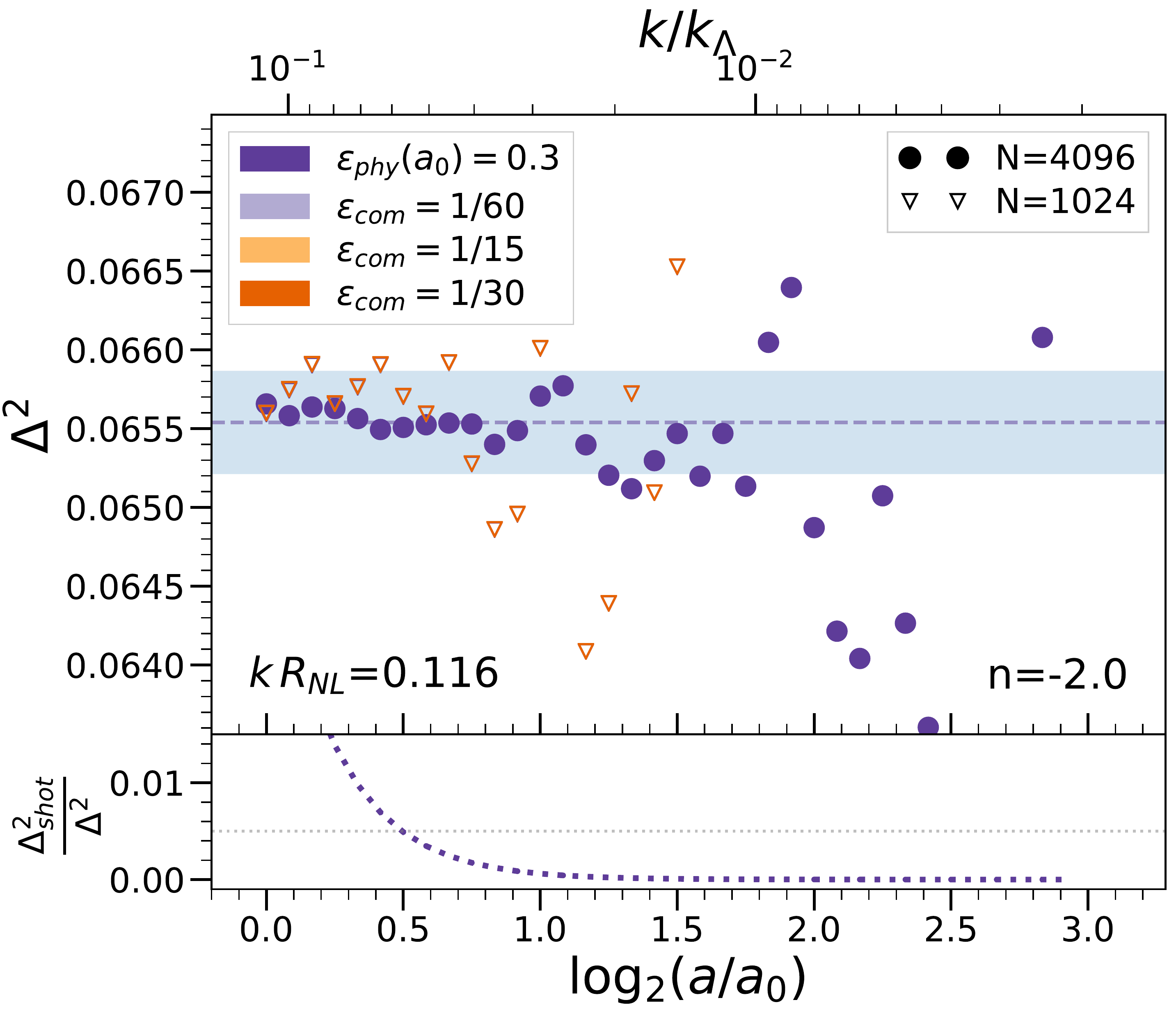}
  \end{subfigure}
  \begin{subfigure}[b]{0.49\textwidth}
    \centering
    \includegraphics[width=\textwidth]{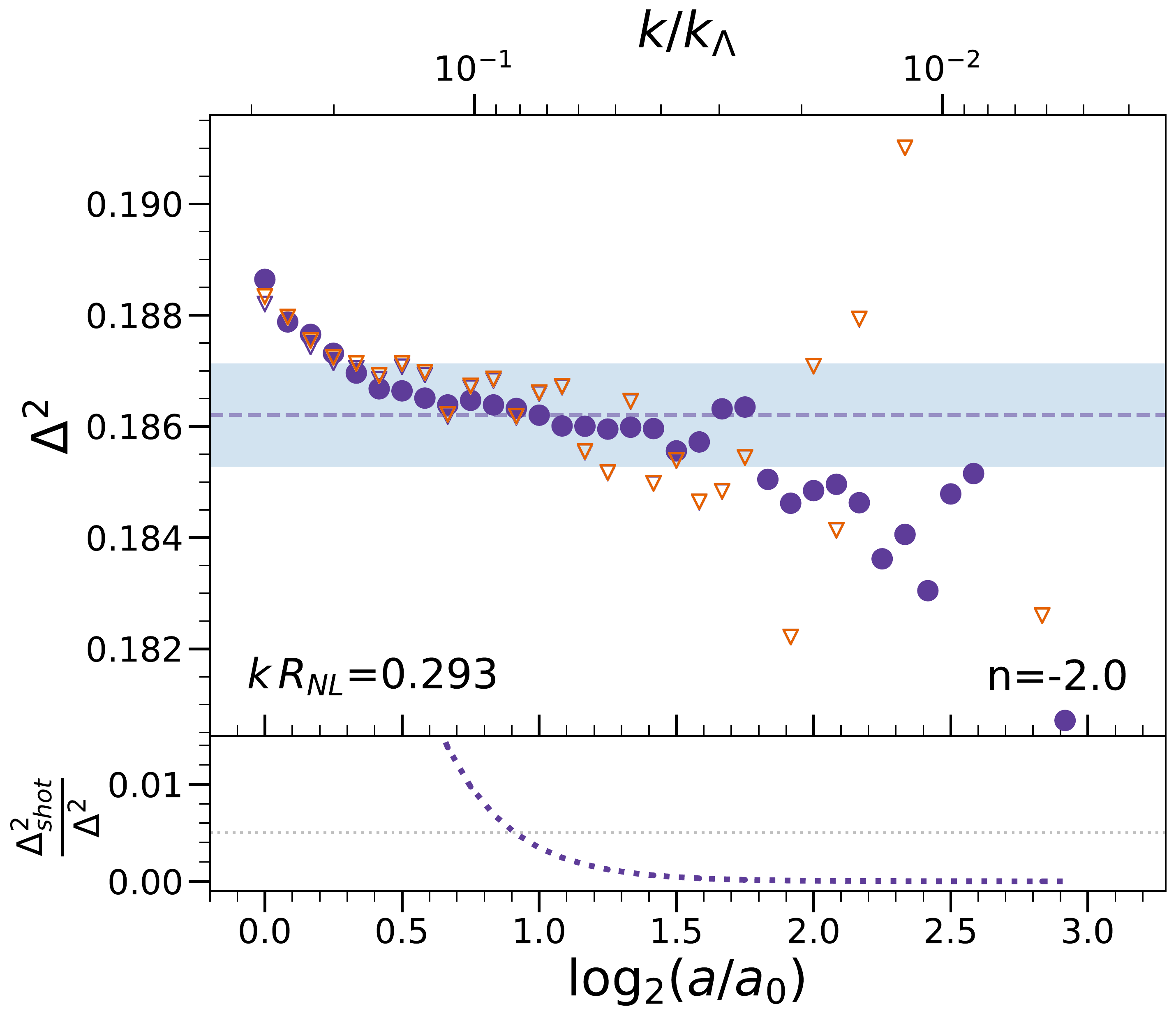}
  \end{subfigure}    
  \medskip
  \begin{subfigure}[b]{0.49\textwidth}
    \centering
    \includegraphics[width=\textwidth]{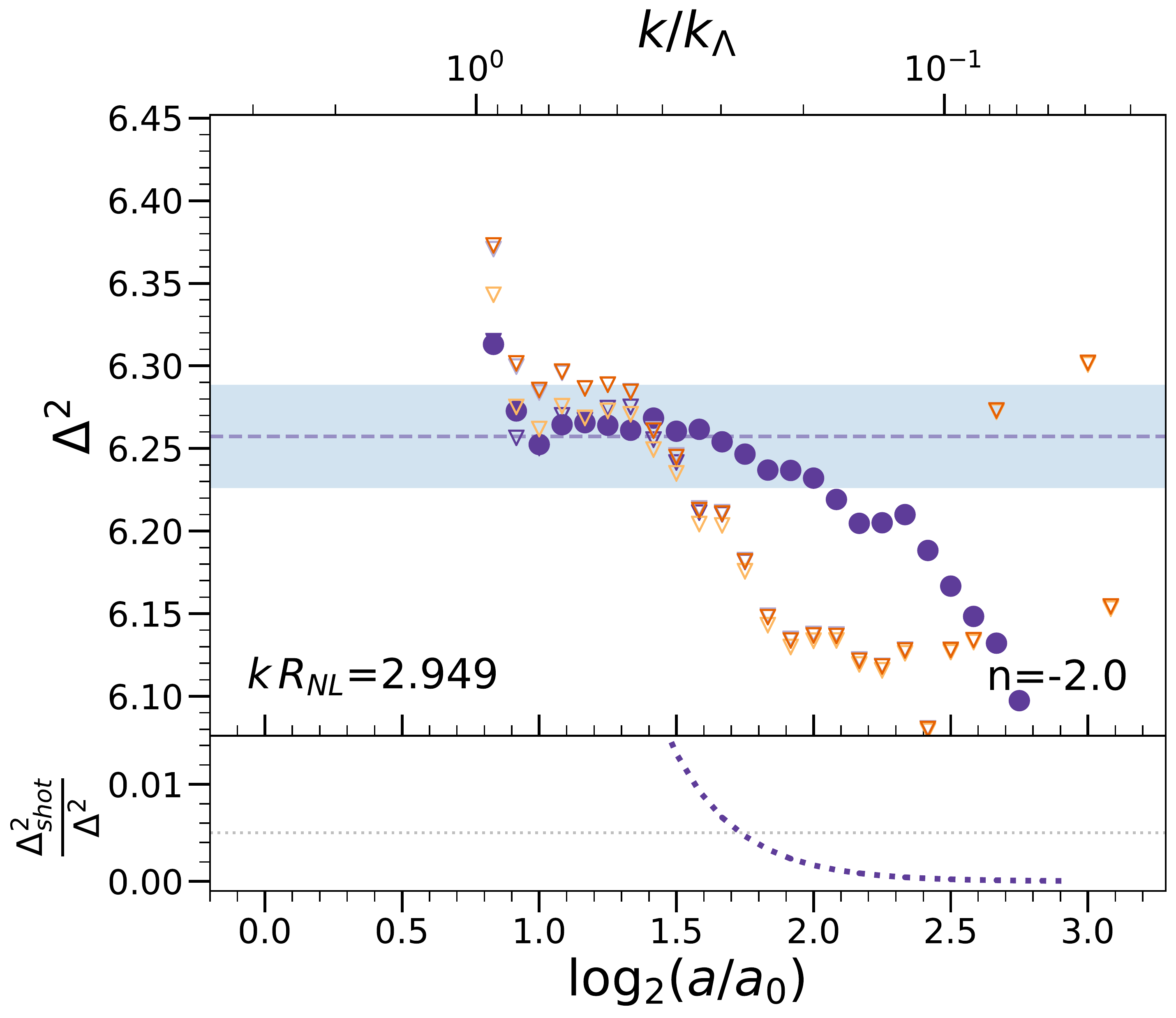}
  \end{subfigure}
  \begin{subfigure}[b]{0.49\textwidth}
    \centering
    \includegraphics[width=\textwidth]{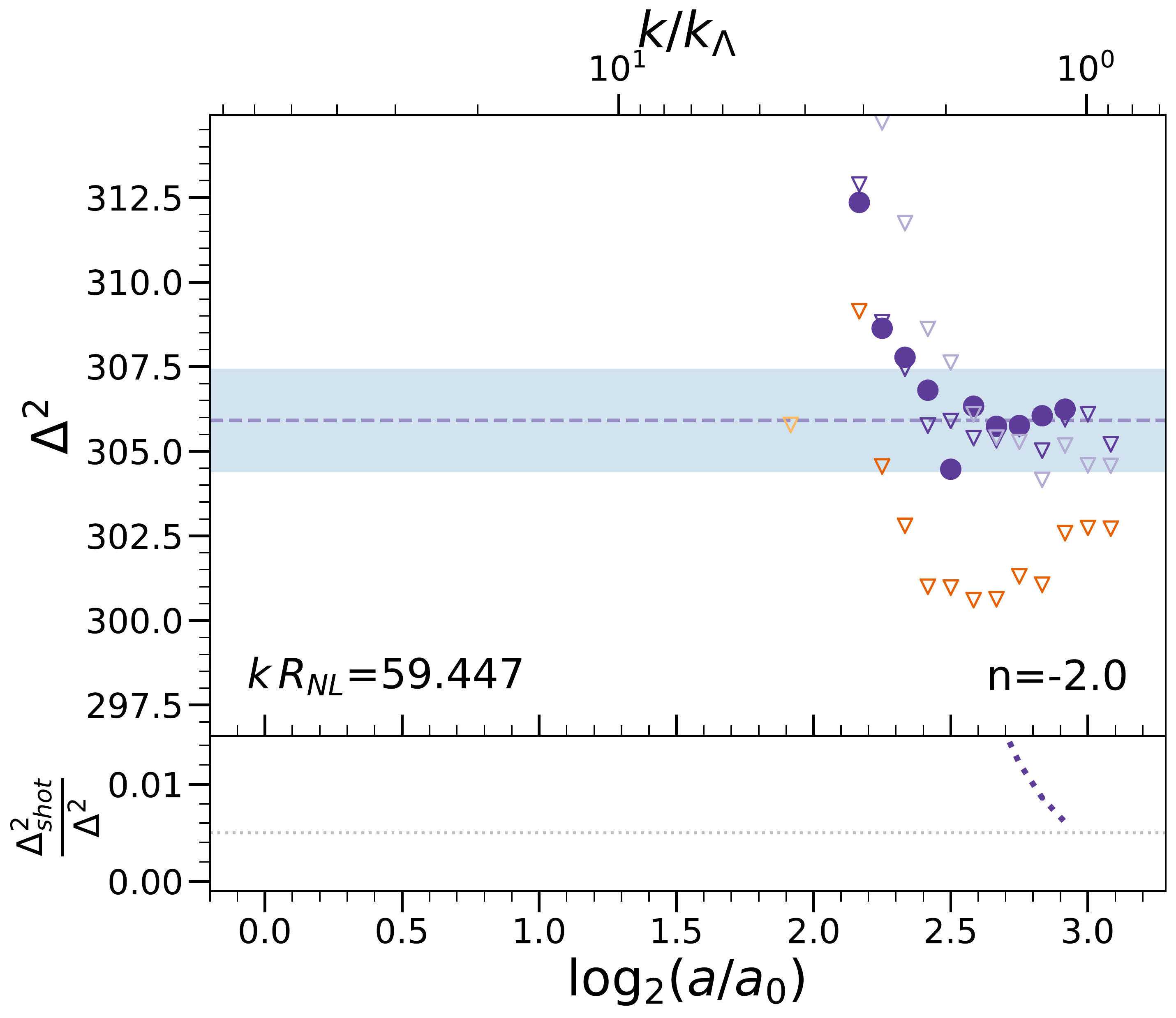}
  \end{subfigure}
  \caption{Same as in \autoref{fig:n15_convplots} but for the five indicated simulations with $n=-2.0$. Circles again correspond to the $N=4096^3$ simulation and triangles to those with $N=1024^3$. Purple corresponds again to physical softening, while the different choices of comoving softening $(1/60, 1/15, 1/30)$ correspond respectively to lilac, yellow and orange. The additional simulations with different softenings allow us to see more clearly its effect on convergence at smaller scales. In particular we note that the chosen physical softening converges as well as the smallest comoving softening, as found in P2 for the 2PCF.}
  \label{fig:n20_convplots}
\end{figure*}

\begin{figure*}
  \begin{subfigure}[b]{0.49\textwidth}
    \centering
    \includegraphics[width=\textwidth]{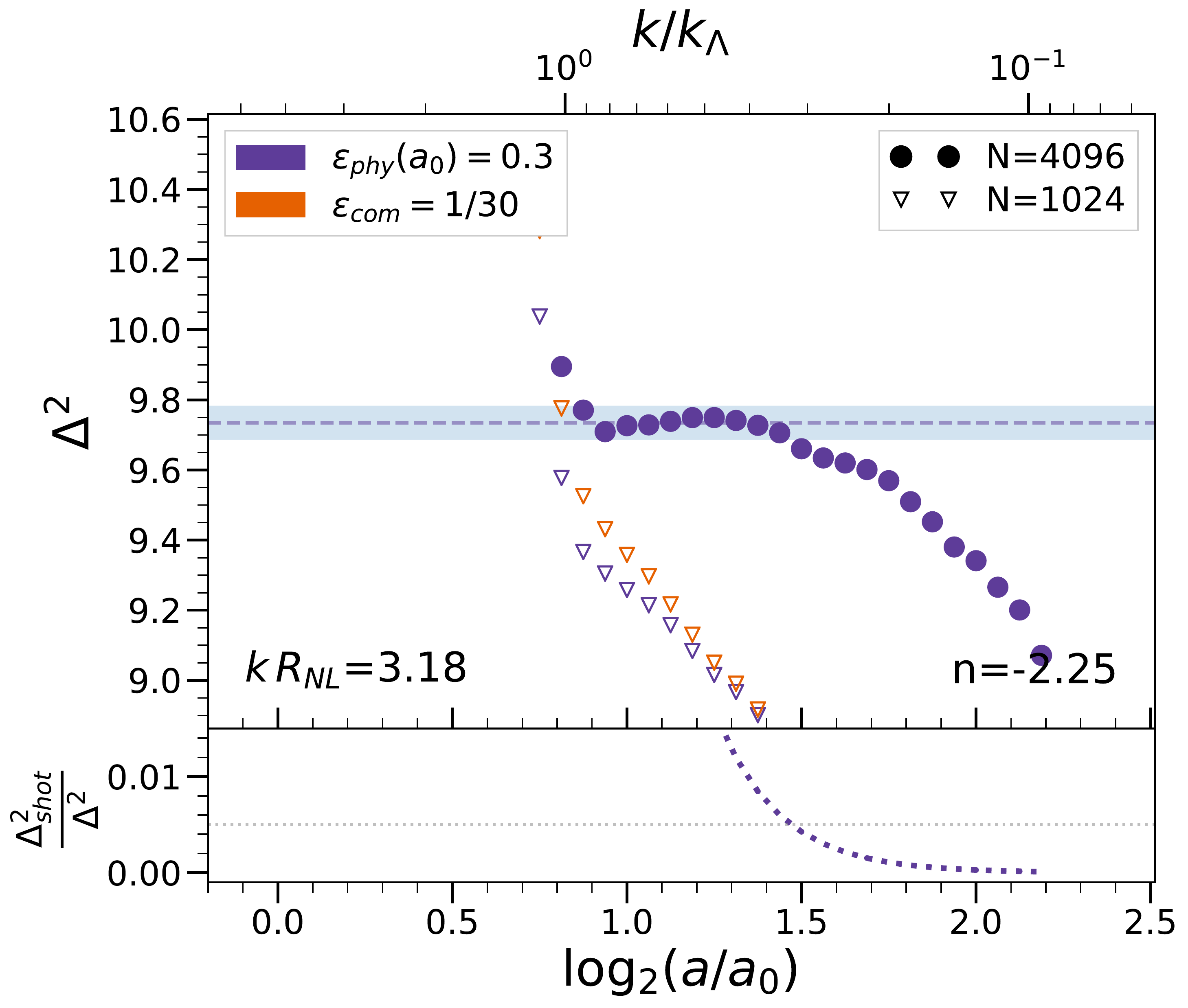}
  \end{subfigure}
  \begin{subfigure}[b]{0.49\textwidth}
    \centering
    \includegraphics[width=\textwidth]{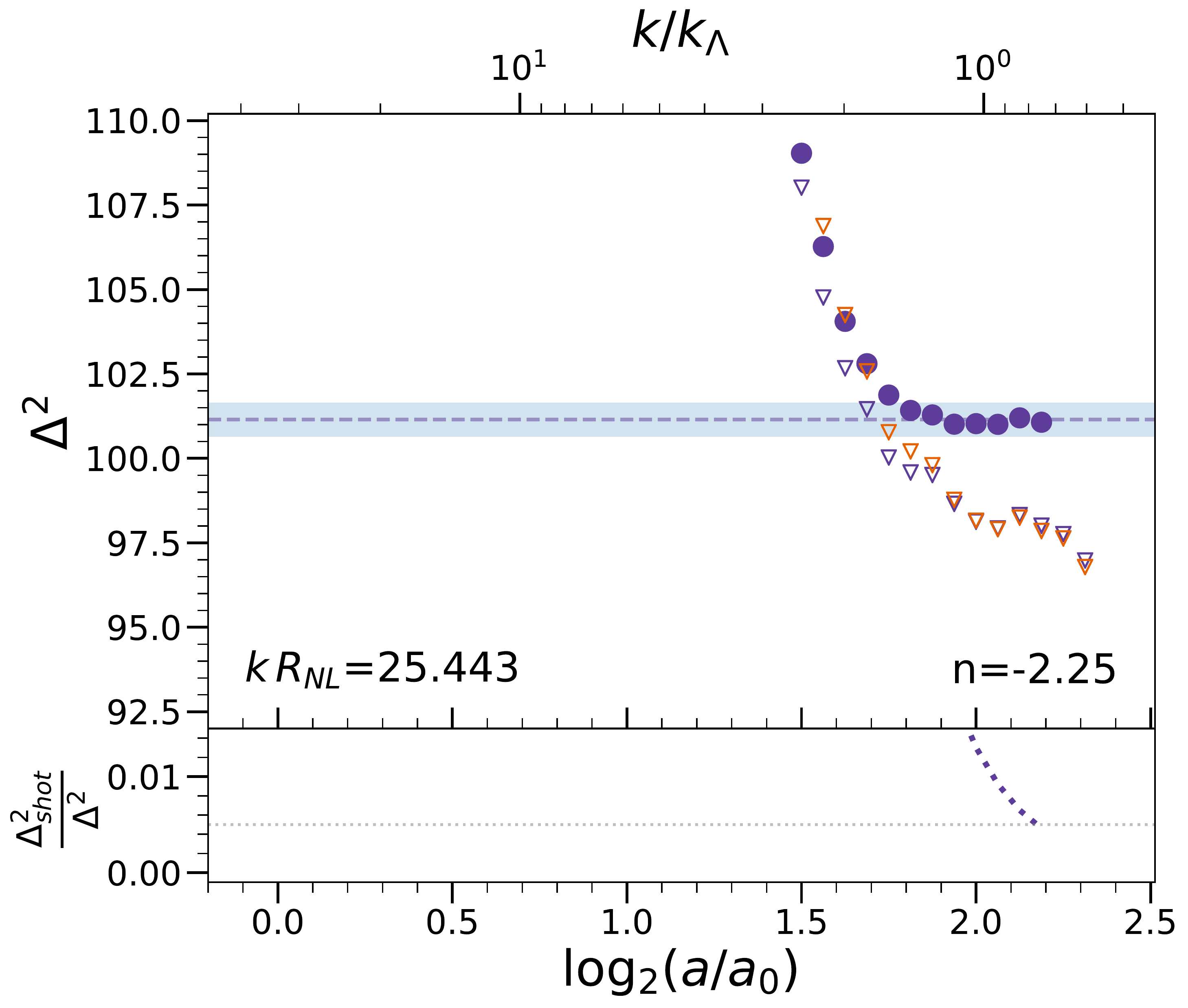}
  \end{subfigure}    
  \caption{Same as in \autoref{fig:n15_convplots} and \autoref{fig:n20_convplots} but for the three indicated simulations with $n=-2.25$. Circles correspond to the $N=4096^3$ simulation and  triangles to $N=1024^3$; physical softening is plotted in purple and orange is used for comoving softening. The light blue region corresponds again to $\pm0.5\%$ of $\Delta^2_{\text{conv}}$, but it appears smaller than on the previous plots because the range on the y-axis has been increased to fit the data in  the smaller simulations. Compared to the plots in the previous two figures, we see clearly how convergence can be obtained only with much larger simulations as $n$ decreases towards $n=-3$.}
  \label{fig:n225_convplots}
\end{figure*}

The accuracy and extent of self-similarity, and how it is limited by the different unphysical simulation parameters, can be better 
visualised by plotting the data in the right panels of \autoref{fig:PS_selfsim} at chosen fixed values of $kR_{NL}$ as a function of 
time, and for the different simulations in \autoref{tab:Simulations}. \autoref{fig:n15_convplots}, \autoref{fig:n20_convplots} and \autoref{fig:n225_convplots} show such 
plots for, respectively, $n=-1.5$, $n=-2.0$ and $n=-2.25$, for the indicated values of $kR_{NL}$.
For convenience we plot also on the upper $x$-axis the corresponding value of $k/k_\Lambda$, and a subplot of the ratio of shot noise over $\Delta^2$.
As $R_{NL}$ is a monotonically growing function of time, $k/k_\Lambda$ \emph{increases from right to left} in each plot: indeed 
$k/k_\Lambda$ can simply be considered as the inverse of the spatial resolution relative to the grid which increases as a function of time.  
Likewise the consecutive plots, at increasing $kR_{NL}$, 
correspond to smaller scales (i.e. larger $k/k_\Lambda$) at any given time. 
The first plot in the first two figures (not plotted for $n=-2.25$) corresponds to the highly linear regime, and a comoving $k$ which is at the first output more than a decade below $k_\Lambda$,
while the last plot for all figures corresponds to the highly non-linear amplitude and a comoving $k$ which is larger than $k_\Lambda$ until close to the last output.

These selected plots illustrate that, in the large range of rescaled wavenumbers and clustering amplitudes shown, we can clearly identify, for each of these three values of $n$, converged values of the PS \emph{at the sub percent level}. The temporal extent and the degree of the convergence vary, but the converged values are indeed independent of the (unphysical) parameters of the different simulations.
Examining the deviations from the self-similar behaviour we can clearly identify the signatures of the three crucial unphysical scales: the initial grid spacing $\Lambda$ (parametrized by $k_\Lambda$), the force softening $\epsilon$ and the box size. 
In the first three panels of \autoref{fig:n15_convplots} and \autoref{fig:n20_convplots}, which show data for
wavenumbers $k \in [0.005,1] k_\Lambda$, we can see that 
the results obtained are insensitive to the gravitational softening (with only some marginally significant differences in the third panel). This is as expected given that the smoothing in all simulations is always much smaller than $\Lambda$.
In these same three panels, on the other hand, the imprint of the two other scales are clearly visible.
Comparing the simulations of the two different box sizes ($1024^3$ and $4096^3$) we can see that the 
larger box improves the self-similarity at larger scales (and later times). While in the first panel this improvement appears just to be due to the reduced noise in the PS estimator (because of the finer sampling in reciprocal space), in the other two bins we observe that the cut-off to the self-similar plateau 
extends to smaller $k/k_\Lambda$. This effect is particularly important in the first panel of \autoref{fig:n225_convplots}, where the smaller simulations ($N=1024^3$) are highly affected by the size of the box and do not reach a converged value. Finally the cut-off (at small scales) to self-similarity clearly depends significantly only on $k/k_{\Lambda}$, and is set by it alone. It is crucial for our extrapolation of our results to non scale-free models that we can thus separate the effects of resolution due to the grid from the effects of finite box size.  

We note that the quality of the convergence to self-similar behaviour is much better for $n=-1.5$ (it is achieved through a larger number of snapshots), because of the finite size effects which become more marked as the spectrum reddens. In particular this effect is crucial in the intermediate scales (typified by the third panel in the first two figures and the first one in \autoref{fig:n225_convplots}) 
where we observe that, for $n=-2.0$, the converged window is still quite small even in the largest
$N=4096^3$ simulation. 

The last panel in these figures, for all three $n$, corresponds to considerably smaller scales and is close to the larger $kR_{NL}$ for which we can identify robustly a converged value (using criteria described in detail below). At these smaller scales the results are now insensitive to the box size, but do show dependence on the softening (except for $n=-2.25$ where box-size effects wipe out this behaviour). The simulations with proper softening and those with the smaller comoving softening show the widest and most coincident regions of self-similarity, while the simulations with larger comoving softening show a suppression of power relative to the self-similar value 
which decreases as $k/k_\Lambda$ does so. Further the same $\epsilon=\Lambda/30$ comoving smoothing shows just marginal ($\sim 1\%$) deviation in the $n=-2.0$ simulation, but more significant ($\sim 3\%$) deviation for $n=-1.5$.
These behaviours are very similar to those which have been analysed and discussed for the 2PCF in P2 for $n=-2.0$. They provide clear evidence that the analogous conclusion may be drawn for the PS as was drawn in P1 and P2 for the 2PCF: the limit on resolution at small scales at any time, corresponding to the largest $k$ at which self-similarity may be attained, is determined by $k_\Lambda$ alone (i.e. by the initial grid spacing, or mass resolution) provided the smoothing is chosen sufficiently small. As can be seen already by comparing the panels of \autoref{fig:n15_convplots}, \autoref{fig:n20_convplots} and \autoref{fig:n225_convplots}, and will be quantified more precisely below, this intrinsic limit on the resolved $k/k_\Lambda$ increases with time (i.e. as a function of  $\log_2(a/a_0)$). Thus there is an upper limit on the required comoving smoothing to attain this intrinsic resolution limit, which depends on the range of scale-factor over which the simulation is run. Our results for $n=-2.0$ indicate, in line with the conclusion of P2 based on the 2PCF, that for a
comoving smoothing below approximately $\Lambda/30$ there is no significant gain in resolution for a typical large LCDM simulation (which has a comparable range of scale-factor, $\log_2(a/a0) \sim  2.5$). For the $n=-1.5$ simulation which runs over a slightly larger range of scale-factor (or a LCDM simulation with a correspondingly higher mass resolution), the limiting comoving smoothing required will be a little smaller.
Likewise for the $n=-2.25$ simulations which span a much lesser range of scale factor, a larger comoving smoothing is sufficient.  
In all cases, on the other hand, as was also observed in P2 for the 2PCF, we see that the chosen proper smoothing --- despite being significantly larger at early times than the comoving smoothing, and thus providing significant numerical economy ---  appears to be quite adequate to attain the intrinsic resolution limit. 

\subsection{Evolution of small scale resolution as a function of time}
\label{Evolution of small scale resolution as a function of time}
We turn now to a more precise quantitative analysis. We focus on the intrinsic upper cut-off (in wavenumber) to resolution determined by $k_\Lambda$, using our data to estimate how it evolves with $a/a_0$.  To do so we first detail the procedure we use to estimate a converged value of $\Delta^2$ in each rescaled bin with an associated error bar (relative to the true physical value). Using these we can then determine, for each comoving $k$, its precision relative to this estimated converged value as a function of time. For any fixed level of precision significantly larger than our intrinsic error on the converged value, we can thus obtain an accurate estimate of the maximal resolved wavenumber at each time.

To estimate converged values for $\Delta^2$ per rescaled bin we proceed as follows\footnote{The exact procedure we use here is slightly different to those used in both P1 and P2.}. Firstly we calculate an estimated converged value (which we denote $\Delta^2_{\text{est}}$) in each rescaled bin as the average PS in the temporal window of fixed width --- here we take six snapshots, but the precise choice is not crucial --- which minimises
\begin{equation}\label{eq:threshold1}
    \frac{\Delta^2_{\text{max}}-\Delta^2_{\text{min}}}{\overline{\Delta^2}}=\alpha\,, 
\end{equation}
where $\Delta^2_{\text{max}}$, $\Delta^2_{\text{min}}$ and
$\overline{\Delta^2}$ are, respectively, the maximum, minimum and average
value in the window. $\alpha$ represents the chosen limit for convergence, fixed at $1\%$ for this study unless stated otherwise, which means that a bin is considered to have converged if $\alpha<0.01$. Finally, we determine for each rescaled bin deemed to have converged under the aforementioned criteria, a (connected) temporal window by seeking the largest one (containing at least three snapshots) with
\begin{equation}\label{eq:conv_criterion}
    \frac{|\Delta^2-\Delta^2_{\text{est}}|}{\Delta^2_{\text{est}}}<\alpha/2
\end{equation}
i.e., values within the ($1\%$) converged region. We denote by $\Delta^2_{\text{conv}}$ the average $\overline{\Delta^2}$ calculated in this largest window, and take it as the value of the dimensionless PS of the bin. The error on this value is estimated as 
\begin{equation}
    \delta(\%) = \pm 0.5\% \sqrt{\frac{w_{\text{min}}}{w}}
\end{equation}
where $w$ is the number of snapshots in the converged window (the one used to calculate $\Delta^2_{\text{conv}}$) and $w_{\text{min}}$ the smallest window for which \autoref{eq:conv_criterion} is fulfilled (by construction $\geq 3$, and
typically, for our parameter choices, $4$ or $5$). 
Although the residual deviations from self-similarity are clearly not uncorrelated between snapshots, visual inspection shows
that this simple choice gives a conservative but reasonable 
error bar that reflects well how confidence in the estimated converged value decreases as the size of the window does.

\begin{figure}
 \centering
 \includegraphics[width=1.1\columnwidth]{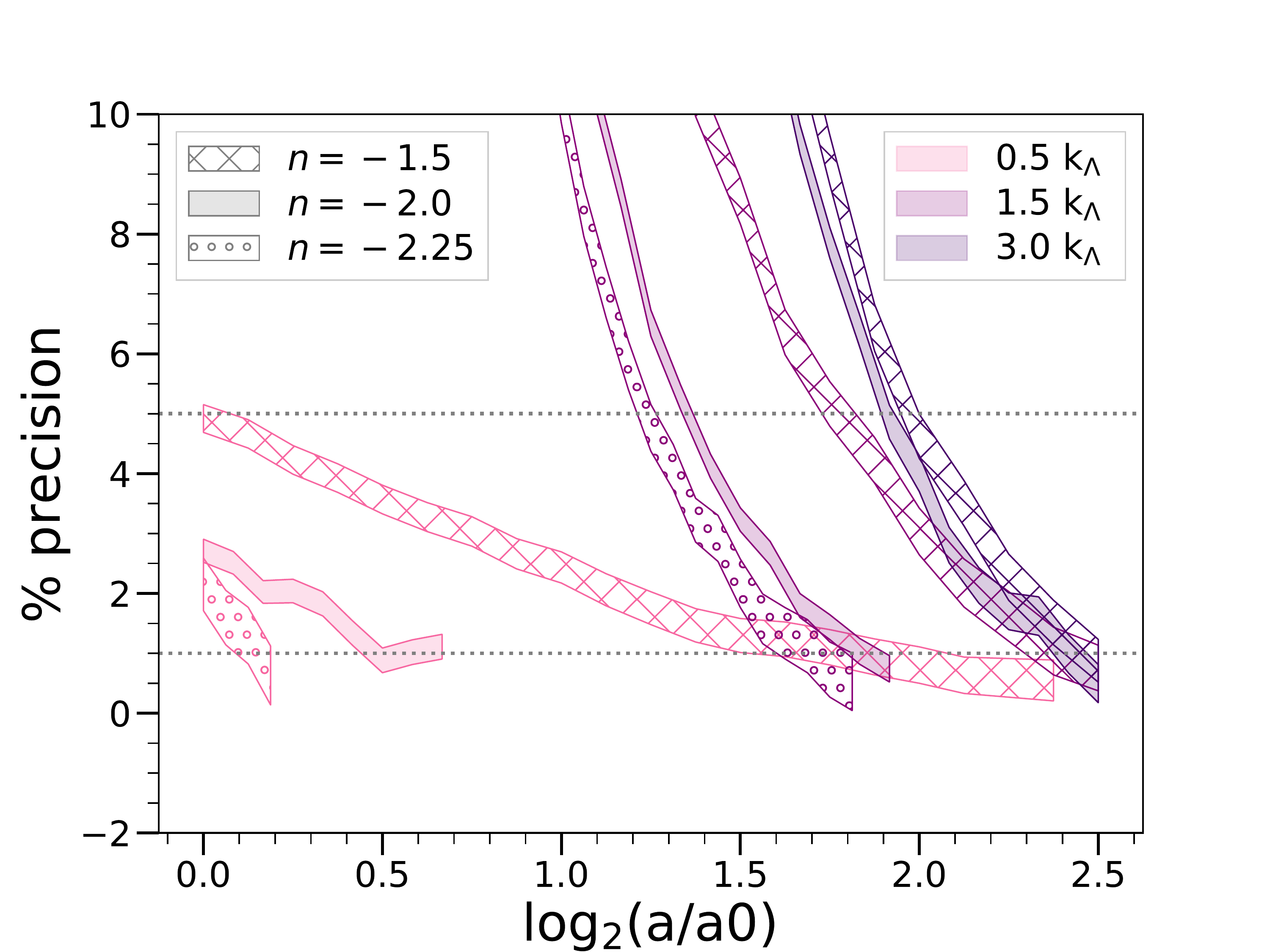}
 \caption{Estimated precision, in percentage, of the PS relative to its true physical value as a function of time, for different selected comoving scales. The indicated confidence intervals correspond to the estimated error $\delta$ in the converged value. The plots in each case are for the largest simulations ($N=4096^3$) for the three spectral indices: $n=-1.5$ (hashed), $n=-2.0$ (filled) and $n=-2.25$ (circles). The horizontal dotted lines correspond to $1\%$ and $5\%$.}
 \label{fig:scale_precision}
\end{figure}

We can now determine the \emph{estimated 
precision} of the PS measured at any scale and time, i.e., the difference between its measured value in a given simulation 
and the true physical value. \autoref{fig:scale_precision} displays the result for five different comoving scales specified
in units of $k_\Lambda$,  $(0.5, 1.5, 3)k_\Lambda$, 
for the $4096^3$ simulations with $n=-1.5$ (hashed), $n=-2.0$ (filled) and $n=-2.25$ (circles).
The horizontal grey dotted lines mark the $5\%$ and $1\%$ 
precision levels.

This plot shows that \emph{the precision of the PS measured at any giving comoving $k$ improves monotonically in time} (at least starting from the time $a=a_0$ at which the first non-linear structures start to form). This is a reflection of the fact that the physical origin of the imprecision probed in this plot is the discretization on the lattice: our plot is restricted to values of $k/k_{\Lambda}$ and time scales where the finite box size effects are negligible, and the data is for the largest simulations (with proper softening) in which the convergence is insensitive to softening. The monotonic improvement of precision is then simply a result of the non-linear gravitational dynamics which efficiently ``transfers power'' from large to small scales i.e. the power at fixed comoving $k$ is determined by power initially at progressively larger scales less affected by the lattice discretization. Correspondingly in a given simulation the maximal wavenumber resolved at any given precision (e.g. $1\%$ or $5\%$) increases monotonically as a function of time.

\begin{figure*}
  \begin{subfigure}[b]{0.49\textwidth}
    \centering
    \includegraphics[width=\textwidth]{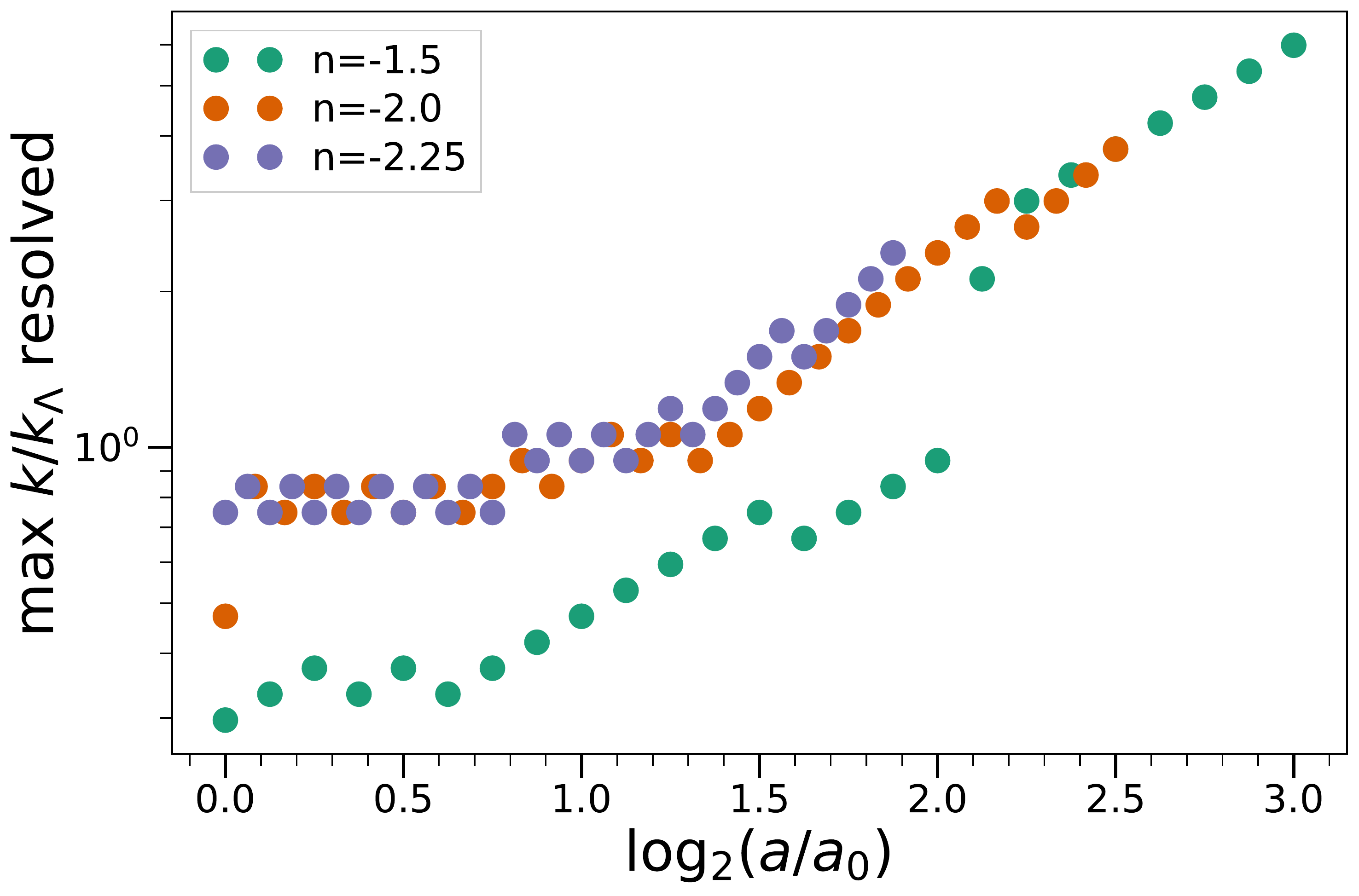}
  \end{subfigure}   
  \begin{subfigure}[b]{0.49\textwidth}
    \centering
    \includegraphics[width=\textwidth]{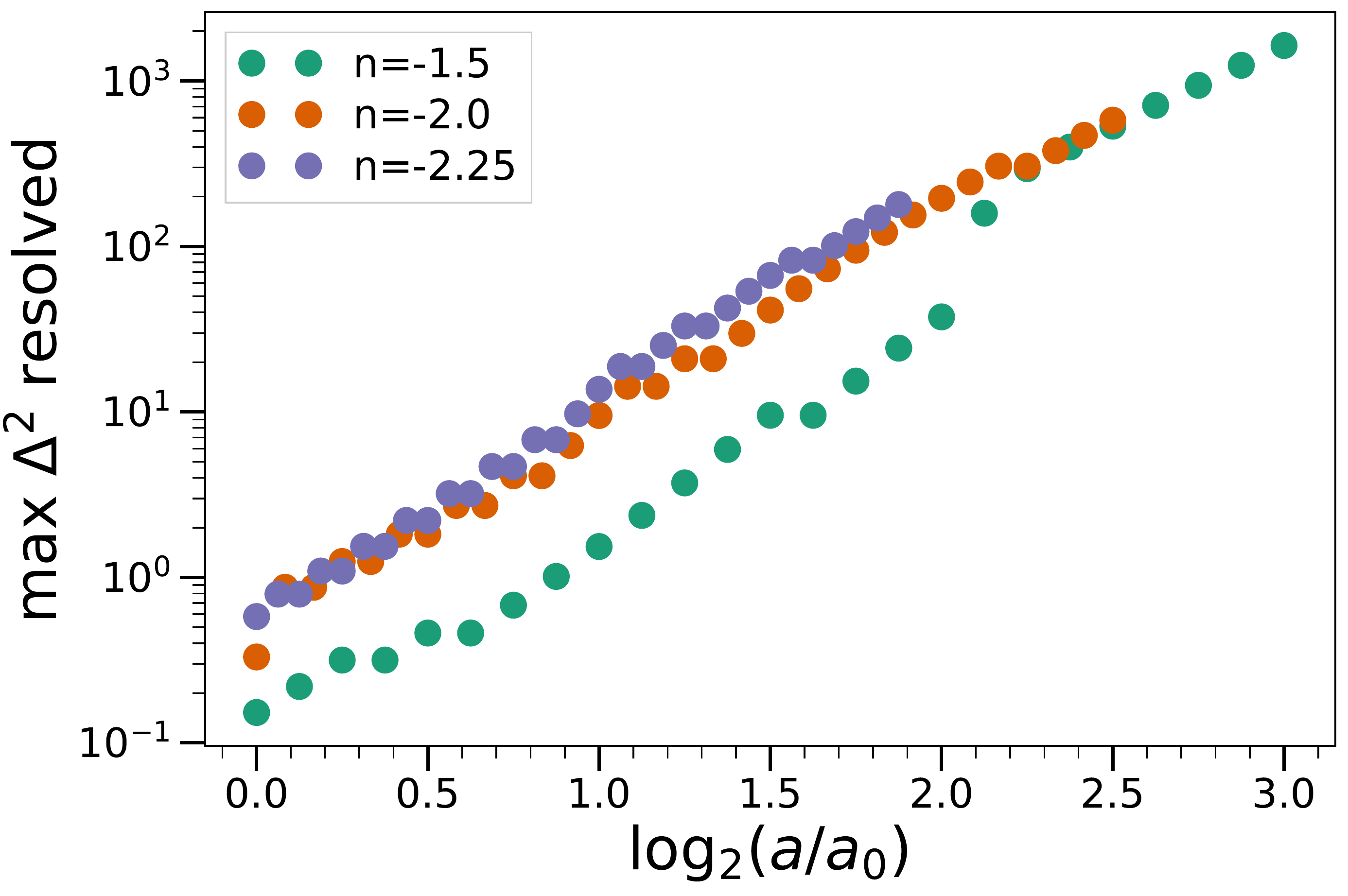}
  \end{subfigure}
  \caption{Maximum resolved wavenumber (in units of the Nyquist frequency of the grid) at $5\%$ precision as a function of time (left panel), and the corresponding maximum resolved dimensionless PS as a function of time (right panel). Results are given for our three different spectral indices ($n=-1.5,-2.0,-2.25$) for the $4096^3$ simulation of each case. We note an approximately $n$-independent resolution in the non-linear regime.
  } 
  \label{fig:resolution}
\end{figure*}

\autoref{fig:scale_precision} also illustrates clearly the qualitative difference between the modes $k<k_\Lambda$ and $k>k_\Lambda$: the former are wavenumbers for which the PS is already resolved well in the initial conditions, while for the latter the physical PS can only be first resolved when the fluctuations at these sub-grid scales are created by non-linear evolution at a level sufficient to dominate over the initial discreteness fluctuations.
For the smallest $k$ the precision of the $n=-2.25$ and $n=-2.0$ simulations, at a given time, are markedly better than for $n=-1.5$, but this difference in precision appears to systematically decrease as $k$ increases,  becoming almost negligible for $k=3k_\Lambda$. The source of the difference for the smaller $k$ arises from the method used to set up initial conditions: as they are defined by a fixed amplitude of $\sigma_{\rm lin}$ at the starting redshift (and thus also at $a=a_0$), a redder spectrum --- for which the mass variance decreases more slowly with scale --- dominates more over the corrections to power from discreteness above $k_\Lambda$. On the other hand, as we are comparing the simulations starting from $a_0$ --- approximately the epoch at which non-linear structure formation first develops in 
them ---  it is not surprising to see that the propagation
of the resolution at non-linear scales may be approximately
model independent when plotted as a function of $a/a_0$: 
it is the evolution of the comoving size of the first virialized structures which might be expected to determine the propagation of resolution, and this can be expected to be primarily determined
by $a/a_0$. 

To probe further this apparent model independence of the resolution as a function of $a/a_0$, 
we show in \autoref{fig:resolution} the evolution of the maximal resolved wavenumber (left panel) and maximal resolved $\Delta^2$ (right panel)  as a function of time. The criterion for resolution is set at a precision of $5\%$ (i.e. the parameter $\alpha=0.05$), which allows us to extend both the temporal range probed and obtain also more constraints from the $n=-2.25$ simulation. We see that indeed, once the non-linear evolution develops sufficiently, starting from approximately $a \sim 4 a_0$, the resolution is approximately the same irrespective of $n$. Further the resolution appears to be correlated with the maximal resolved $\Delta^2$, consistent with the hypothesis that it can essentially be determined in a model independent manner.  

\subsection{Shot noise and discreteness effects}\label{sec:PoisonNoise}

As discussed in \autoref{sec:PScalculation}, it is 
interesting to consider whether shot noise can be  used to model deviations from the physical PS due to $N$-body discretization. In our analysis above we did not subtract a Poisson noise term from the calculated PS other than to account for that induced by random down-sampling of our larger samples.
Indeed our method is designed to pick up any
dependence on any unphysical scale (and $\Lambda$ in particular) as a break from self-similarity. It 
does not require that we make any assumption on the functional form of such modifications or the 
range of applicable scales. We now consider what our
results above tell us specifically about 
this question.

First from \autoref{fig:PS_selfsim} we see that 
the PS does approach the shot noise level at asymptotically large $k$ (as it must,  by definition). However it can be seen that it only attains this behaviour at scales $k\gg k_{\Lambda}$, significantly larger than those where we have seen the simulations to be converged (as shown in \autoref{fig:scale_precision}). At early times
in particular it can be seen that the measured PS is in fact below the shot noise level. Subtracting such a term clearly cannot then give an approximation for the (positive) physical PS, and it would evidently spoil the apparent self-similarity of our data (seen in the right panels) in a range of scale. At intermediate times, especially for the smaller $n=-2.25$ index, we can observe that the simulations depart from self-similarity at k smaller than those where the shot noise dominates the PS, indicating that the reason for this self-similarity breaking is not due to a shot noise of this form. Only in the latest snapshot of our simulations can we see that this term dominates at the scales close to those
where there are deviations from self-similarity.

These conclusions can be seen more fully and further quantified from the lower subplots of \autoref{fig:n15_convplots}, \autoref{fig:n20_convplots} and \autoref{fig:n225_convplots}. We can compare directly the deviation of the PS actually 
observed in the simulations from its converged value, with the fraction represented by 
a shot noise term (shown in the lower subpanels). 
In all cases we see that, in some part of the range where the PS falls within $0.5\%$ of its estimated converged value, the shot noise represents a significantly larger fraction. Shot noise thus 
overestimates the deviations of the PS.
This makes it clear that if such a term is subtracted, it will in fact degrade the convergence
in some range i.e. the PS with subtracted shot noise will approximate the physical PS with a reduced range of scale and/or time.  

\begin{figure}
 \centering
 \includegraphics[width=1.1\columnwidth]{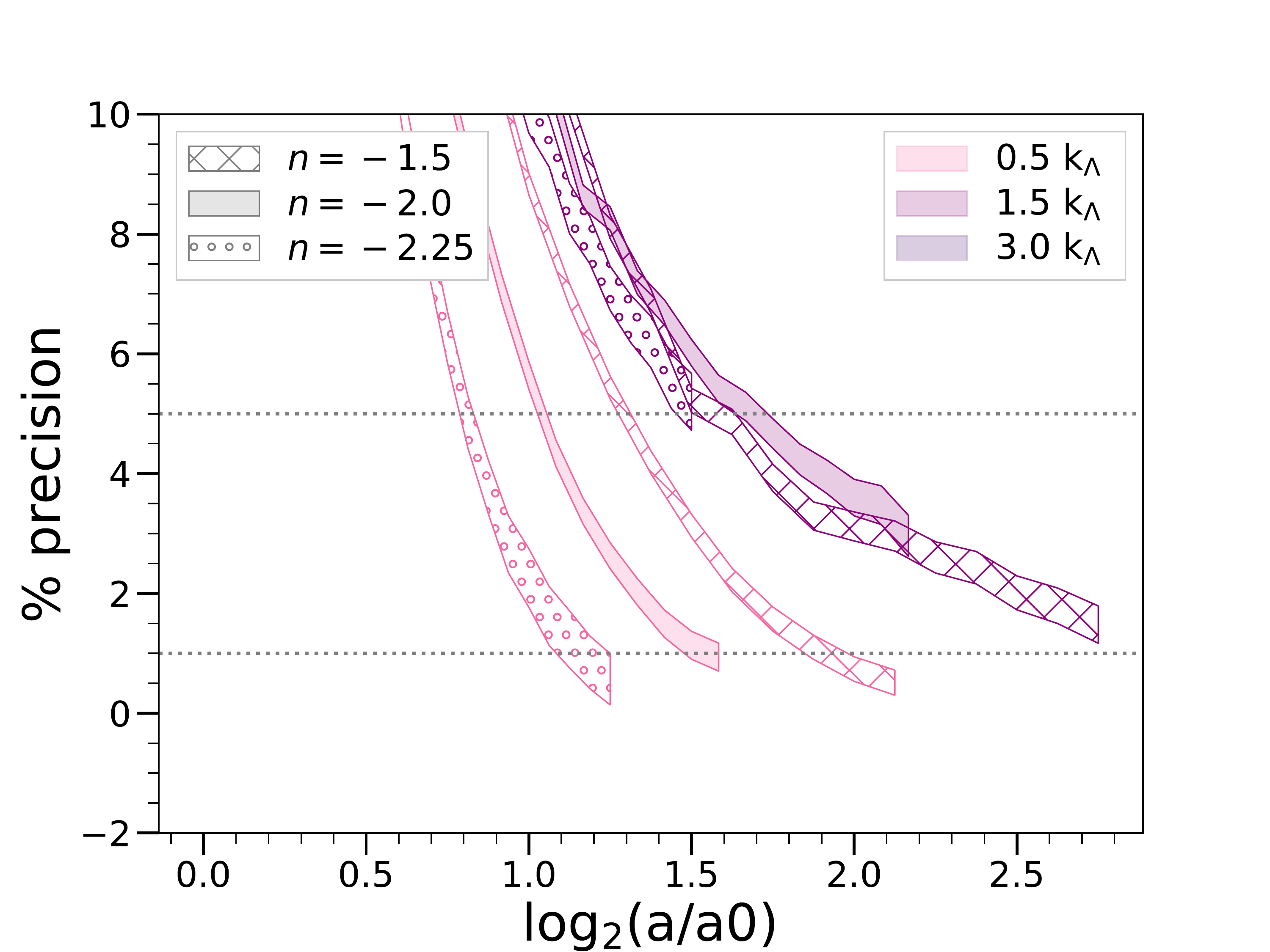}
 \caption{Equivalent analysis to \autoref{fig:scale_precision} where a shot noise term ($P(k)=\Lambda^3$) has been subtracted from the computed PS. It shows the estimated precision, in percentage, of the PS relative to its true physical value as a function of time, for different selected comoving scales. The indicated confidence intervals correspond to the estimated error $\delta$ in the converged value. The plots in each case are for the largest simulations ($N=4096^3$) for the three spectral indices: $n=-1.5$ (hashed), $n=-2.0$ (filled) and $n=-2.25$ (circles). The horizontal dotted lines correspond to $1\%$ and $5\%$.}
 \label{fig:PoisonNoise}
\end{figure}

To see how significant this undesired effect of shot noise subtraction may be, we have  performed an identical analysis as the one described in the subsections above, but now subtracting a Poisson noise term. \autoref{fig:PoisonNoise} is the equivalent of \autoref{fig:scale_precision} with the subtraction. We see clearly that the convergence is indeed destroyed at early times, as anticipated above.
We also see that convergence for larger k is affected, and  $1\%$ precision for $k>k_{\Lambda}$ is never really achieved, and we do not resolve $k\sim k_{\Lambda}$ even at $5\%$ and later times. 

In conclusion, and under our self-similar analysis of our simulations, we understand that a term of such form does not appropriately describe the shape of PS. While the smallest scales per snapshots are indeed well dominated by $L^3/N$, we argue that subtracting a term of this form at all times and scales is not justified, and does indeed degrade the PS resolution. The region where this procedure would be reasonable $(k\gg k_{\Lambda})$ is not accessible at a $1\%$ precision for comparable simulations, and thus the procedure becomes redundant.

\section{Resolution limits for non scale-free cosmologies}
While our method, by construction, is limited strictly to
scale-free cosmologies, our underlying motivation is of course to quantify the resolution in simulations of non scale-free cosmologies such as LCDM or variants of it. Such cosmologies are not
really so very different from scale-free cosmologies for what concerns their non-linear evolution: for this purpose their
PS can be considered to be an adiabatic interpolation of power-law spectra, with the modified expansion rate due to dark energy only coming into play at very low redshift. 

\begin{figure}
 \centering
 \includegraphics[width=1.\columnwidth]{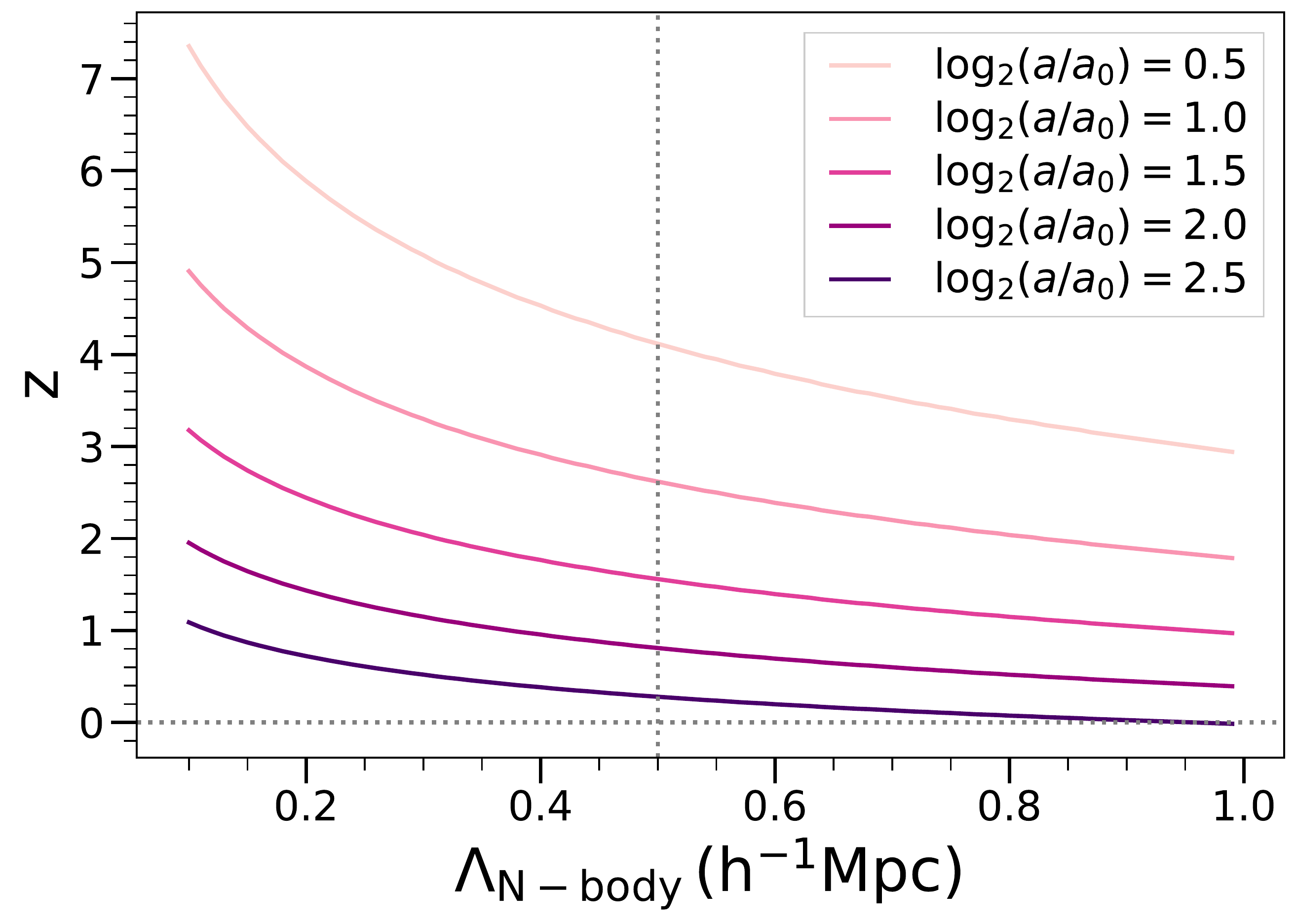}
 \caption{Redshift $z$ corresponding to different fixed values of $\log_2(a/a_0)$ as a function of mean interparticle spacing $\Lambda$, using a standard LCDM cosmology (``Planck 2013'', \protect\cite{planck_2013}). As discussed in the text, combining this plot with the curves for in \autoref{fig:scale_precision} and \autoref{fig:resolution} we can infer a conservative bound on attainable precision as a function of redshift in an LCDM simulation.}
 \label{fig:zeff}
\end{figure}

We have anticipated this extrapolation of our results above by choosing to simulate, as far as practicable, scale-free models with $n$ in the range relevant to LCDM, and focusing on the dependence of our results on these exponents. Further we have characterised how resolution depends on time in terms of a scale factor relative to $a_0$ which can be defined, as given in \autoref{eq:def_a0}, for any cosmology, and has the simple physical meaning as the time when non-linear structures start to develop.  \autoref{fig:zeff} illustrates how the parameter 
$\log_2(a/a_0)$ maps to the redshift in a simulation of a standard LCDM model (``Planck 2013'', \cite{planck_2013}). The mapping is just a function of the mean interparticle spacing 
$\Lambda$ (and the linear power spectrum of the model) as these allow the determination of $a_0$. This means that, given a simulation of a determined grid spacing $\Lambda$, one can always find a one-to-one relation between the desired evolved redshift of the LCDM and our time variable $\log_{2}(a/a_0)$. As discussed in P1, non-EdS expansion at low redshift introduces the possibility of mapping the time rather than the scale-factor, but the difference in the effective $\log_2(a/a_0)$ is in practice very small, and we will neglect it here.

\begin{figure}
 \centering
 \includegraphics[width=1.\columnwidth]{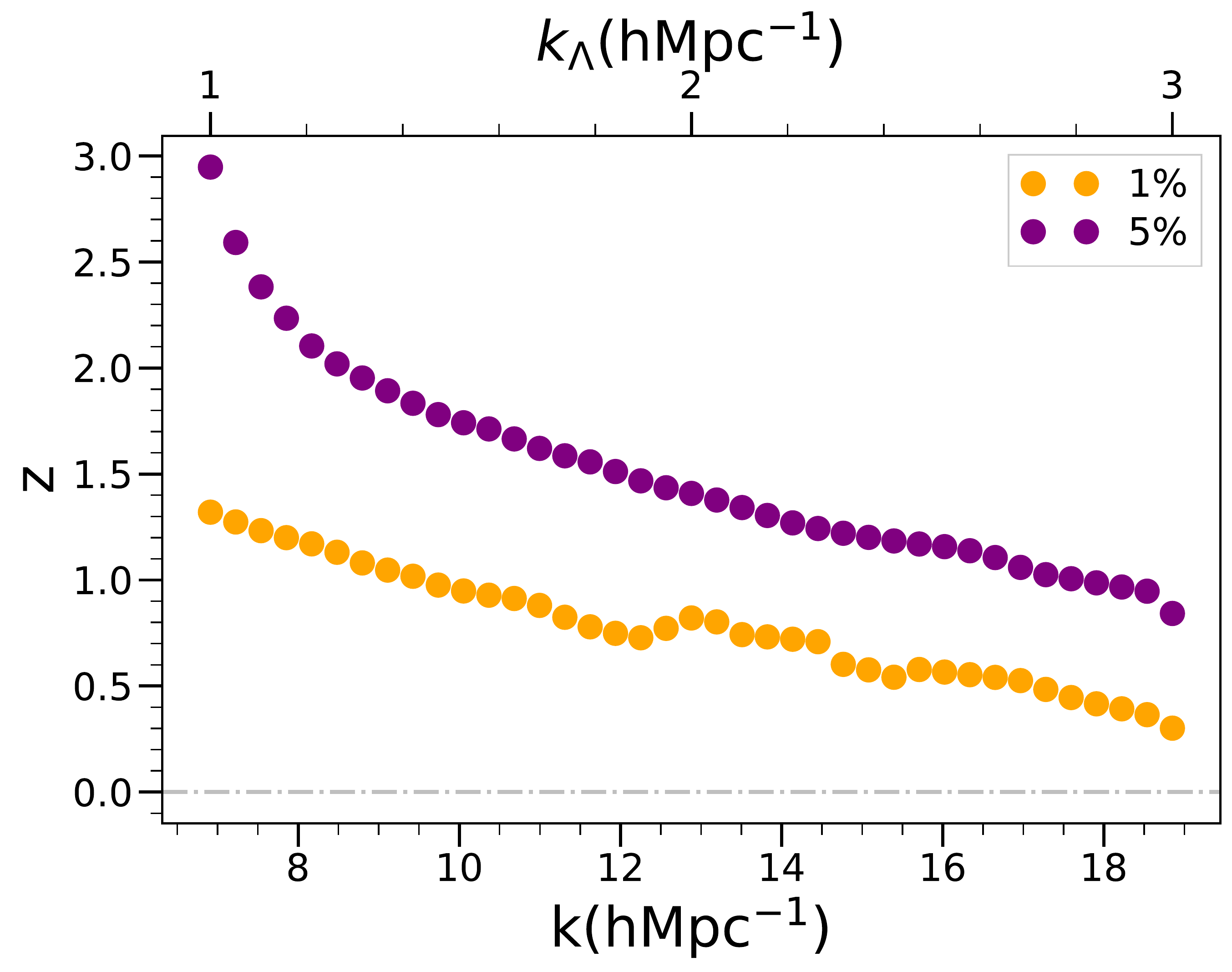}
 \caption{Resolved k as a function of redshift for a fixed value of the mean interparticle spacing $\Lambda=0.5h^{-1}\text{Mpc}$, using a standard LCDM cosmology (``Planck 2013'', \protect\cite{planck_2013}). Orange points correspond to $1\%$ precision, $\alpha=0.01$, while purple points represent a $5\%$ precision, $\alpha=0.05$. We've added an axis for the $k_{\Lambda}$ of the hypothetical LCDM simulation for an easier extrapolation with \autoref{fig:zeff} and \autoref{fig:scale_precision}}
 \label{fig:LCDMcomparison}
\end{figure}

Our analysis above shows that we have two quite different regimes for the evolution of small scale resolution. For the modes $k > k_\Lambda$, which are not modelled in the initial conditions and which always describe the strongly non-linear regime, resolution is approximately $n$ independent 
as a function of $\log_2(a/a_0)$. For these modes it seems very reasonable that we can carry over straightforwardly our results from the scale-free case to the general case (modulo small possible corrections due to non EdS evolution). That is, given the grid spacing of the simulation the results in \autoref{fig:scale_precision} (or \autoref{fig:resolution}) can simply be converted to resolution as a function of redshift using \autoref{fig:zeff}. As an example, \autoref{fig:LCDMcomparison} shows, for a model $\Lambda=0.5h^{-1}\text{Mpc}$ simulation, the smallest scale which we will have access to as a function of redshift at a $1\%$ and $5\%$ precision.  

On the other hand, for the modes $k < k_\Lambda$ we have seen that the resolution as a function of time shows $n$-dependence. These differences can be understood as arising directly from the initial conditions, and are thus essentially dependent on the behaviour with scale
of the mass variance around the initial grid spacing. For a generic LCDM type cosmological model with a slowly varying exponent we can thus expect to determine the resolution from that of a scale-free model which, at scales $\sim\Lambda$,  approximates the behaviour of the variance. Such an effective exponent can be estimated as: 
\begin{equation}
    n_{\text{eff}}=-3-2\left.\frac{d\log\sigma}{d\log R}\right|_{R=\Lambda}
\end{equation}

Taking the aforementioned LCDM model ("Planck 2013"), its estimated $n_{\text{eff}}$ is situated typically around or below our smallest simulated exponent, at least for $\Lambda$s of typical (current) N-body simulations, $n\sim(-2.5,-2)$ for $\Lambda\in[0.1,1]$. The resolution of $n=-2.25$ shows the same tendency as $n=-2.0$, inferred from \autoref{fig:scale_precision}. We thus conclude that we can use our results for $n=-2.0$ to obtain conservative bounds on precision for numerical simulations of such models.

These statements assume of course implicitly that the method of 
setting up initial conditions is like that used in our simulations, that 
convergence  has been established with respect to time-stepping and any all other numerical parameters, and that the box size is sufficiently large so that finite box effects are negligible at these scales considered.

\section{Discussion and Conclusions}
We conclude by discussing our results first in relation to the closely related recent studies in P1 and P2, and then in relation to the broader literature on the power spectrum of dark matter in the non-linear regime. 

Our analysis here is a development of that already presented in  
P1 and P2, which focused on the self-similarity of the 2PCF, and also in \citet{Leroy2020}, which explored the self-similarity of halo mass and 2PCF functions. It further confirms that the study of self-similarity in scale-free cosmologies, using the methods introduced in P1, 
is a very powerful tool to quantify resolution of cosmological $N$-body simulations. We have performed a complementary analysis to the real-space analysis of the 2PCF in P1, giving a precise quantitative characterisation of the evolution of resolution of the PS in k-space. Doing so we have tested even further the accuracy of the \Abacus code by demonstrating  that it can measure this essential cosmological statistic to an accuracy well below percent level in a wide range of scales. While these previous studies used suites of simulations for a single power-law ($n=-2.0$) and a single box size ($N=1024^3$), we have considered a suite for both different power-laws and also a much larger box ($N=4096^3$). This has allowed us to identify the effects of finite box size, and separate them clearly --- over a range which depends strongly on the exponent $n$ --- from the resolution limits due to the finite particle density and gravitational smoothing. 
This is essential for robustly extrapolating these results on small scale resolution limits to non scale-free models. 

It is interesting to compare our conclusions concerning the resolution of the PS to those obtained for the 2PCF in P1 and P2.
Qualitatively, our analysis simply confirms that the main conclusions in these works map into reciprocal space just as one would naively expect. In particular P1 and P2 concluded that the lower cut-off scale to resolution for the 2PCF is, provided gravitational force smoothing is sufficiently small, fixed by the initial grid spacing. Further this scale is found, starting from a certain time, to be a monotonically decreasing function of time. We have drawn here identical conclusions about the PS in k-space. Looked at in more detail qualitatively, and quantitatively, the results in the two spaces are not, however, related in the simple manner one might expect. For example, the lower cut-off to resolution in real space was found, at the few percent level, to decrease in proportion to $a^{-1/2}$ starting from $0.2 \Lambda$. 
Mapped into reciprocal space this would give a resolved
$k$ very considerably larger than what we have found (and with a very different $a$ dependence). Comparing any of our various measurements confirms the conclusion that the resolution in reciprocal space is much poorer than in direct space, in the simple sense that inverse wavenumber characterising it is much smaller than $\pi$ divided by the spatial resolution. Thus, in reciprocal space, discreteness effects 
are ``smeared" out over a larger range of scale than in direct space, where they are more localized. The same is true also of the effects of gravitational smoothing: indeed we have found here that for the $n=-2.0$ simulation a comoving smoothing of $\Lambda/30$ produces few 
percent  ($1\%$) deviations from self-similarity at $k \sim 2 k_\Lambda$ (i.e. $k \sim \pi/(15 \epsilon)$) 
while in real space similar deviations are seen at only $\sim \Lambda/10$ (i.e. $r \sim 3 \epsilon$).

As we have discussed in the introduction the question of resolution of different statistics, and in particular the 2PCF or PS, as measured in $N$-body simulations, has long been a subject of discussion in the literature. In the light of new and more accurate surveys it has become an even more important practical issue. Most current studies are based on the comparison of results from simulations at different mass resolution, where the highest available resolution is considered as the ``true" converged value
relative to which precision is measured. Our analysis here is based on the much more robust identification of converged values possible in scale-free simulations. As we have discussed in the last section
these results can be extrapolated in a simple manner to obtain robust conservative resolution limits in LCDM models.

Laying aside the numerous caveats that we have underlined should be born in mind when extrapolating to any other set of simulations (and notably different codes), it is interesting to compare our conclusions with typical assumptions about resolution of PS measurements in the literature. In the recent literature on simulations of LCDM cosmologies focused on extracting very accurate measurements at intermediate scales, such assumptions are generally quite consistent with, and indeed often somewhat more conservative than, the limits we have found.
For example, \citet{Ishiyama2020} assume that $2\%$ accuracy is attained for $k < k_\Lambda/2$ at $z=0$, \citet{Heitmann2020} a $5\%$ accuracy at the same range up to $z<1$, or 
\citet{Angulo2020} assume a $1\%$ accuracy up to the same scale.  In studies which extend their analysis 
of the PS to the strongly non-linear regime, on the other hand, it is commonplace to include wavenumbers $k$ approaching $k \sim \pi/\epsilon$, 
for $\epsilon \ll \Lambda$ (e.g. \citet{smith2003stable, springel_etal_2018}). 
Indeed strong clustering develops down to these small scales, and it may be that it provides a 
reasonable approximation to the true physical
limit, allowing at the very least qualitative conclusions that are correct. Nevertheless our analysis indicates that resolution in fact appears to degrade very rapidly for $k$ significantly larger than a few $k_\Lambda$. This suggests (as argued previously in e.g. \cite{splinter_1998, romeo_etal_2008, discreteness3_mjbm} that caution should be exercised in making use of simulation data in the range of $k \gg k_{\Lambda}$.

\begin{figure}
 \centering
 \includegraphics[width=1.1\columnwidth]{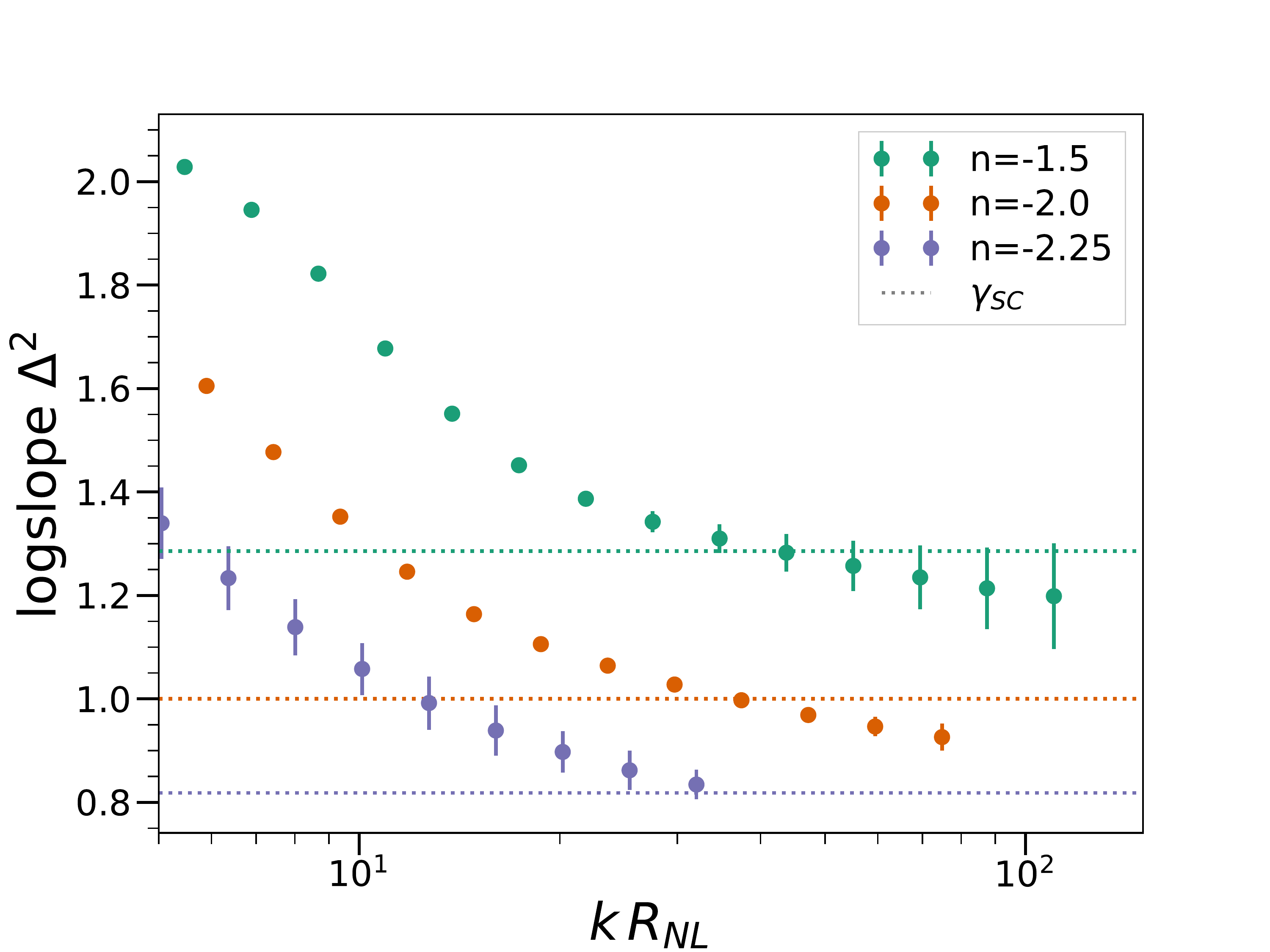}
 \caption{Estimated converged logarithmic slopes of the PS for each of the three simulated values of $n$ in our biggest $4096^3$ simulations. The dashed horizontal lines indicate the exponent corresponding to stable clustering, $\gamma=3(3+n)/(5+n)$.
 The estimated precision of these measurements and the derivation of the error bars are detailed in the text. The curves are remarkably close to consistency with the hypothesis of asymptotic stable clustering, with at most, given our indicated error bars, marginal evidence for slightly lower values. We note that the studies by \protect\cite{smith2003stable} and \protect\cite{widrow_etal2009} estimate  asymptotic converged slopes equal to $0.91$ for $n=-1.5$, $0.77$ for $n=-2.0$ and $0.7$ for $n=-2.25$.
 }
 \label{fig:logslopes}
\end{figure}

An illustration of this conclusion is given by
considering what has been inferred from
$N$-body simulations about the behaviour of 
the PS at asymptotically large $k$. In early work Peebles
\citep{peebles_1974} envisaged the possibility that the non-linear structures would decouple as they shrink in comoving coordinates, the so-called ``stable clustering" hypothesis. Simulations of scale-free models have been an ideal testing ground for it, as Peebles derived in this case a simple analytical prediction for a power-law behaviour corresponding to $\Delta^2(k) \sim k^\gamma$ where 
$\gamma=3(3+n)/(5+n)$. While early simulation studies \citep{efstathiou1988gravitational, colombi_etal_1996, bertschinger_98,jain+bertschinger_1998, valageas_etal_2000} showed evidence for its validity  --- at least for the shallower  indices in which the finite box effects were not overwhelming for smaller simulations --- studies by  \cite{smith2003stable} and \cite{widrow_etal2009} concluded that it breaks down, and further that 
the power spectrum's behaviour at small scales can be characterized by a single, but significantly smaller, $n$-dependent exponent. 
The robustness of these conclusions have been questioned in \citet{benhaiem2013self, benhaiem_etal_2017} using a new analysis of simulations of the same size ($N=256^3$) as those of \cite{smith2003stable}. Shown in 
\autoref{fig:logslopes} are the estimated converged values of the logarithmic slope of $\Delta^2$, from  our $4096^3$ simulation of each of the three exponents. These have been estimated by applying to the logarithmic slope the same method described in detail in Section \ref{sec:NumSim} for the PS, with parameter $\alpha=0.05$. This corresponds to an estimated precision of at most $\pm 2.5\%$, i.e. about twice the diameter of the plotted points. The error bars shown in the plot correspond to
$\pm$ the absolute value of the biggest difference in the converged values obtained in our different simulations (using identical convergence criteria), and indicate a significantly larger systematic error (towards an underestimate of the exponent)
due to smoothing in the last few bins. 
These results show  measured exponents remarkably close to the predicted stable clustering values, with at most marginal evidence for a deviation from stable clustering. The considerably smaller
exponents derived in the studies of \cite{smith2003stable} and \cite{widrow_etal2009} are clearly due to the extrapolation to 
poorly resolved scales (\cite{smith2003stable} used $\epsilon=\Lambda/15$ and $N=256^3$). We will present elsewhere further detailed study of this issue, including also in particular the corresponding real space analysis of the 2PCF which can provide a consistency check on any possible detection of an asymptotic exponent.

\section*{Acknowledgements} 
S.M thanks the Institute for Theory and Computation (ITC) for hosting her in early 2020 and acknowledges the Fondation CFM pour la Recherche for financial support.
D.J.E. is supported by the U.S. Department of Energy grant DE-SC0013178 and as a Simons Foundation Investigator. L.H.G. is also supported by the Simons Foundation.  This research used resources of the Oak Ridge Leadership Computing Facility at the Oak Ridge National Laboratory, which is supported by the Office of Science of the U.S. Department of Energy under Contract No. DE-AC05-00OR22725.
The \textsc{AbacusSummit} simulations have been supported by OLCF projects AST135 and AST145, the latter through the Department of Energy ALCC program.

\section*{Data availability} 
The data reported in the article are provided in csv format in the online supplementary material.




\bibliographystyle{mnras}
\bibliography{Bibliography} 




\appendix

\section{Details of Power Spectrum calculation}
\label{appendix}
For mass assignment we use the Triangular Shape Cloud (TSC) assignment method \citep{Book} given by
\begin{equation}\label{eq:TSC}
    W(x_i)=\begin{cases} 0.75 - x_i^2 & \mbox{if } |x_i|<0.5 \\ \dfrac{(1.5-|x_i|)^2}{2} & \mbox{if } 0.5<|x_i|<1.5 \\ 0 & \mbox{else} \end{cases}
\end{equation}

 After performing the Fourier transform using the publicly available code pyfftw \citep{pyfftw}, we correct 
by dividing  by the ``alias" sum of the Fourier transform of the window function 
 $P^f \rightarrow P^f/ \sum_n W^2(\mathbf{k}+2k_g\mathbf{n})$ \citep[e.g.,][]{Angulo2007, Sato2011, Takahashi2012}, which is given exactly for the TSC assignement by
\begin{equation}
    \sum_n W^2(\mathbf{k}+2k_g\mathbf{n})\approx\prod_j\left[1-\sin^2\left(\frac{\pi\mathbf{k}_j}{2k_{g}}\right)+\frac{2}{15}\sin^4\left(\frac{\pi\mathbf{k}_j}{2k_{g}}\right)\right].
\end{equation}
where the product is over the three components of  $\mathbf{k}$.


\bsp	
\label{lastpage}
\end{document}